\documentclass[12pt,preprint]{aastex}

\lefthead{K. Krisciunas {\em et al.}}
\righthead{Photometry of Nearby Supernovae}

\begin{document} 

\title{Optical and Infrared Photometry of the Nearby Type Ia 
Supernovae 1999ee, 2000bh, 2000ca, and 2001ba}

\author{Kevin Krisciunas,$^{1,2}$
Mark M. Phillips,$^1$
Nicholas B. Suntzeff,$^2$
S. E. Persson,$^3$ 
Mario Hamuy,$^3$
Roberto Antezana,$^5$
Pablo Candia,$^2$
Alejandro Clocchiatti,$^8$
Darren L. DePoy,$^{12}$
Lisa M. Germany,$^4$
Luis Gonzalez,$^5$
Sergio Gonzalez,$^2$
Wojtek Krzeminski,$^1$
Jos\'{e} Maza,$^5$
Peter E. Nugent,$^9$
Yulei Qiu,$^6$
Armin Rest,$^2$
Miguel Roth,$^1$ 
Maximilian Stritzinger,$^{10}$
L.-G. Strolger,$^7$
Ian Thompson,$^3$
T. B. Williams,$^{11}$ and
Marina Wischnjewsky$^{5,13}$}
\affil{$^1$Las Campanas Observatory, Carnegie Observatories, Casilla 601, La Serena, Chile \\
$^2$Cerro Tololo Inter-American Observatory, National Optical Astronomy \\ 
Observatory,\altaffilmark{14} Casilla 603, La Serena, Chile \\
$^3$Observatories of the Carnegie Institution of Washington, 813 Santa Barbara St.,
Pasadena, CA 91101-1292 \\
$^4$European Southern Observatory, Casilla 19001, Santiago, Chile \\
$^5$Departmento de Astronom\'{i}a, Universidad de Chile, Casilla 36-D, Santiago, Chile \\
$^6$ National Astronomical Observatories, Chinese Academy of
Sciences,  Beijing 100012 \\
$^7$Space Telescope Science Institute, 3700 San Martin Drive, Baltimore, MD 21218 \\
$^8$Univ. Cat\'{o}lica de Chile, Astronom\'{i}a y Astrofisica, Casilla 104, Santiago 22, Chile \\
$^9$Lawrence Berkeley Laboratory, 1 Cyclotron Road, MS 50-F, Berkeley, California 94720 \\
$^{10}$Max-Planck-Institut f\"{u}r Astrophysik, Karl-Schwarzschild-Str.\ 1, 85741 Garching, Germany \\
$^{11}$Rutgers University, Dept. of Physics and Astronomy, 136 Frelinghuysen Rd., Piscataway, NJ 08855-0849 \\
$^{12}$Department of Astronomy, The Ohio State University, 140 West 18th Avenue, Columbus, OH 43210 \\
$^{13}$Deceased}
\altaffiltext{14}{The National Optical Astronomy
Observatory is operated by the Association of Universities for
Research in Astronomy, Inc., under cooperative agreement with the
National Science Foundation.}

\email {kkrisciunas, arest, nsuntzeff@noao.edu \\
sgh, pcandia@ctiosz.ctio.noao.edu \\
mmp, wojtek, miguel@lco.cl  \\
aclocchi@astro.puc.cl \\
depoy@astronomy.ohio-state.edu \\
lgermany@eso.org \\ 
penugent@lbl.gov \\
rantezan, lgonzale, jmaza@das.uchile.cl \\
qiuyl@bao.ac.cn \\
stritzin@mpa-garching.mpg.de \\
strolger@stsci.edu \\
mhamuy, persson, ian@ociw.edu \\
williams@physics.rutgers.edu }

\begin{abstract}

We present near infrared photometry of the Type Ia supernova 1999ee;  also, optical 
and infrared photometry of the Type Ia SNe 2000bh, 2000ca, and 2001ba.  For SNe
1999ee and 2000bh we present the first-ever SN photometry at 1.035 $\mu$m (the
$Y$-band).  We present K-corrections which transform the infrared photometry in
the observer's frame to the supernova rest frame.  Using our infrared
K-corrections and stretch factors derived from optical photometry, we construct
$JHK$ templates which can be used to determine the apparent magnitudes at
maximum if one has some data in the window $-$12 to +10 d with respect to
T($B_{max}$). Following up previous work on the uniformity of $V \; minus$ IR
loci of Type Ia supernovae of mid-range decline rates, we present unreddened
loci for slow decliners.  We also discuss evidence for a {\em continuous}
change of color at a given epoch as a function of decline rate. 

\end{abstract}

\parindent = 0 mm


\section{Introduction}

\parindent = 9 mm

Type Ia supernovae (SNe) are well known to the {\em cognoscenti} as
standardizable candles for determining extragalactic distances.  They are
considered to be the explosions of CO white dwarfs in binary systems.  Using
optical light curves, Phillips (1993), Hamuy et al. (1996a), and Phillips et
al. (1999)  established the relationship between the absolute magnitudes at
maximum and the decline rates of these SNe.  The characteristic parameter is
$\Delta$m$_{15}$($B$), the number of magnitudes that the SN declines in the
first 15 days after $B$-band maximum.\footnote[15]{The method actually uses
the $B$-, $V$- and $I$-band light curves to obtain a characteristic value of
$\Delta$m$_{15}$($B$).} Riess, Press, \& Kirshner (1996), Riess et al.
(1998), and Jha, Riess, \& Kirshner (2004) have elaborated three versions of
the Multicolor Light Curve Shape (MLCS) method. This method now correlates
the light curve shapes in $UBVRI$ with the absolute magnitudes.  In the MLCS
method the characteristic parameter is $\Delta$, the number of $V$-band
magnitudes that a Type Ia SN at maximum is brighter than ($\Delta < 0$), or
fainter than ($\Delta > 0$) a fiducial object of M$_V$(max) = $-$19.47 on a
scale where H$_0$ = 63 km s$^{-1}$ Mpc$^{-1}$ . Finally, the stretch method
of Perlmutter et al. (1997) is used to stretch or shrink $B$- and $V$-band
light curves in the time axis to match a fiducial.\footnote[16]{The stretch
method does not work with $R$-and $I$-band light curves, except around 
T($B_{max}$), because of the shoulder in the $R$-band
light curves and the secondary maximum in the $I$-band ones.  
See Fig. \ref{i_stretch} below.  For roughly 90
percent of Type Ia SNe, the strength of the secondary hump is correlated with
$\Delta$m$_{15}$($B$) (Krisciunas et al. 2001), hence with the absolute
magnitudes at maximum.} Since brighter SNe are more slowly declining, the
stretch factor is less than 1.0 for the intrinsically brighter SNe, and
greater than 1.0 for the intrinsically fainter ones.  Typically,
$\Delta$m$_{15}$($B$) = 1.1 corresponds to $\Delta$ = 0 and stretch factor
$s$ = 1.0.

While the optical light curves of most Type Ia SNe fit into some kind of an
ordering scheme, with the marked increase of SN discoveries in the past few
years, more and more objects are being found which do not fit the patterns.  
SN~1999ac was a slow riser, fast decliner (Labbe et al. 2001, Phillips et
al. 2003).  SN~2000cx was a fast riser, slow decliner and had unusually blue
$B-V$ colors after T($B_{max}$) + 30 d (Li et al. 2001, Candia et al.
2003).  Recently, Li et al. (2003) describe the even stranger SN~2002cx,
which had premaximum spectra like SN 1991T (the classical hot Type Ia SN), a
luminosity like SN 1991bg (the classic sub-luminous object), a slow
late-time decline, and unidentified spectral lines. These
``pathological'' examples should motivate breakthroughs in the modeling of
Type Ia SNe, allowing us to understand better the more normal objects.

Following the pioneering papers of Elias et al. (1981, 1985) and the
photometry of SN~1986G (Frogel et al. 1987), essentially no infrared
light curves of Type Ia SNe were obtained until the appearance of
SN~1998bu.
(See Meikle 2000 for a compilation of all the IR photometry
available three years ago.)  In 1999 two of us (MMP and NBS) started
regular observing campaigns at Las Campanas Observatory and Cerro
Tololo Inter-American Observatory to obtain optical and infrared
light curves of supernovae, primarily Type Ia's.  For this we
have used mainly the LCO 1-m Swope telescope and the CTIO
1-m Yale-AURA-Lisbon-Ohio (YALO) telescope. 

In this paper we report optical and infrared photometry of the Type Ia
SNe 2000bh, 2000ca, and 2001ba.  We present infrared photometry of
SN~1999ee. Previous extensive $UBVRIz$ optical photometry of SN~1999ee 
has been published by Stritzinger et al. (2002).

A significant fraction of the data reported in this paper was obtained for
the Supernova Optical and Infrared Survey (SOIRS), a project organized and
carried out in 1999 and 2000 (Hamuy principal investigator).  See Hamuy et al.
(2001) for further details.  SOIRS also included a photographic search for
supernovae carried out at Cerro El Roble (Maza 1979). Amongst SOIRS
discoveries are three of the four SNe reported here.  We note that the
optical and IR photometry and spectroscopy of SN~1999ee (one of the prime
SOIRS targets) is the largest such dataset ever obtained for a Type Ia SN.

\section{Observations}

\subsection{Photometric Calibration}

Supernova light curves can be derived from images taken on photometric and
non-photometric nights thanks to the use of digital array
detectors.  This works providing that calibration has been done on some
photometric nights, and it assumes that the exposures are long
enough for any variations of transparency of the sky covered by the
digital array to even out.  In our experience this works better for optical
photometry than infrared photometry.  In the latter the field of view
is usually quite small, and one takes many short exposures while
dithering the position of the telescope.  If the sky is
non-photometric, only the center of the subsequently constructed
mosaic is useable.  Thus, one hopes to have an appropriately bright
field star or two close to the SN so that data from all nights can be
utilized.

One of our concerns is whether observing through clouds affects the
photometric colors: are clouds {\em grey}? Serkowski (1970) demonstrated that
$U-B$, $B-V$, and $V-R$ colors were affected less than 0.01 mag even though
he was observing through up to eight magnitudes of cloud cover.  Walker
et al. (1971) and Olsen (1983) reported similar reassuring results.

As is common, we have tied the optical photometry of our SNe to the
$UBVRI$ standards of Landolt (1992a).  This allows us to calibrate
the {\em field stars} near the SNe.  However, the spectra of SNe are
unlike spectra of normal stars, so ideally one would then devise a
set of corrections to place the SN photometry on a particular
photometric system, such as that of Bessell (1990).  Stritzinger et
al. (2002) attempted this with their extensive optical photometry of
SN 1999ee, but in the end decided not to apply any corrections,
citing ``rather disappointing'' results.  By contrast, Krisciunas et al.
(2003) were much more encouraged with their similar endeavors with SN
2001el.  We used optical data obtained with the CTIO 1.5-m telescope,
the CTIO 0.9-m telescope, and the YALO 1-m telescope.  We were able
to devise corrections for the $B$- and $V$-bands which placed the
photometry on the Bessell filter system, tightened up the light
curves, and eliminated a known source of systematic error in the
$B-V$ light curves.  However, if we had applied our derived
corrections for the $R$- and $I$-bands, the light curves derived from
imagery obtained on different telescopes would have actually been
pulled further apart, not tightened up.  Hence, such corrections were not
applied to the $R$ and $I$ photometry.

For the calibration of the photometry presented here we used a combination
of aperture photometry within the {\sc iraf}\footnote[17]
{{\sc iraf} is distributed by the National Optical Astronomy
Observatory, which is operated by the Association of Universities
for Research in Astronomy, Inc., under cooperative agreement with
the National Science Foundation.} environment and analysis using 
point spread function ({\sc psf}) magnitudes obtained with either
{\sc daophot} (Stetson 1987, 1990) or John Tonry's version of {\sc
vista} (Terndrup, Lauer, \& Stover 1984).  Field star magnitudes were 
typically determined via
aperture photometry, while instrumental magnitudes of the SNe were
{\sc psf} magnitudes.  When necessary we performed image subtraction
using a script written by Brian Schmidt and based on the algorithm
of Alard \& Lupton (1998).

Much of the IR photometry presented in this paper was obtained from
imagery obtained with the LCO 1-m and LCO 2.5-m telescopes.
The LCO 1-m infrared camera utilizes a
Rockwell 256 $\times$ 256 NICMOS-3 HgCdTe array, giving a plate
scale of 0.60 arcsec pixel$^{-1}$.  The LCO 2.5-m data for SN~1999ee
were obtained with the {\sc cirsi} camera, giving a plate scale of
0.20 arcsec pixel$^{-1}$.  The LCO 2.5-m data for SNe 2000bh, 2000ca,
and 2001ba were obtained with a camera containing a similar Rockwell
array to that in the 1.0-m camera, but with a plate scale of 0.35 arcsec 
pixel$^{-1}$.

A majority of the $J$- and $H$-band of SN~1999ee was obtained with
the optical and infrared camera {\sc andicam} on the YALO 1-m telescope.
The infrared camera contained a 1024 $\times$ 1024 HgCdTe array by Rockwell,
giving a plate scale of 0.22 arcsec pixel$^{-1}$.  The optical camera in
{\sc andicam}, which was used for the SN~2001ba observations presented
here, utilized a Loral 2048 $\times$ 2048 CCD, giving a plate scale of
0.30 arcsec pixel$^{-1}$.

It was with the LCO 1-m telescope that Persson et al. (1998) established their
system of IR standards.  Thus, no corrections need to be made to our LCO
photometry to place it on the Persson system.  For the reduction of the YALO
photometry to the Persson system we used color terms derived from observations of
Persson standards made in February 2000.  For YALO $J$ the color term is $-$0.043
$\pm$ 0.005, scaling $J-H$.  For YALO $H$ the color term is +0.015 $\pm$ 0.005,
scaling $J-H$.  For YALO $K$ the color term is $-$0.003 $\pm$ 0.005, scaling
$J-K$.  For the case of our YALO IR photometry of SN~1999ee, we also used spectra
of SN~1999ee (Hamuy et al. 2002) and appropriate filter profiles to correct the
YALO photometry for the non-stellar spectral energy distribution of the SN.

In the case of $z$-band photometry, Hamuy (2001, Appendix B)  
gives synthetic $z$-band magnitudes of 20 spectrophotometric standards (Stone
\& Baldwin 1983), quoting uncertainties of $\pm$ 0.02 mag.  For more information
on the $z$-band filter see Schneider, Gunn, \& Hoessel (1983).  We made
$UBVRIz$ observations of the SN~2000bh and 2000ca fields from 29
January to 1 February 2003, along with some of these
spectrophotometric standards and Landolt (1992a) fields, to
calibrate the $z$-band photometry of the SNe.  (Additional $BVRI$
calibration was obtained from images of May 2000, when these two
SNe were active.)  Details of the $z$-band calibration are given in
Appendix A.

The $Y$-band at 1.035 $\mu$m is a new photometric band whose
prime advantages are described by Hillenbrand et al. (2002).  
It is a relatively clean atmospheric window.  $Y$-band photometry
should be relatively independent of the specific IR detector,
relatively insensitive to differences in transparency vs. wavelength and 
nightly variations of water vapor at a given site.  For a tabulation
and graphical representation of the $Y$-band filter transmission, 
see Table 1 and Fig. 1, respectively, of Hillenbrand et al. (2002).

If we derive $Y$-band values for Vega, Sirius, and the Sun (see
Appendix A of Krisciunas et al. 2003), we  obtain the following
linear transformation:

\begin {equation}
(Y-K_s) \; = \; -0.013 \; + \; 1.614 \; (J_s-K_s) \; ,
\end {equation}

\parindent = 0 mm

where $J_s$ (``J-short'') and $K_s$ (``K-short'') are the
equivalent magnitudes in the system of Persson et al. (1998).  
Many of the Persson et al. standards have $J_s-K_s$ colors in
between those of Vega and the Sun, so we may rely on interpolation
for the presumed $Y-K_s$ colors.  Of various Persson et al. stars
observed by us in the $Y$-band, we note that P9183 has $Y$ =
12.459, according to Eq. 1, while Hillenbrand et al. (2002) give
$Y$ = 12.296 $\pm$ 0.030, which is 0.16 mag (5.4-$\sigma$) brighter
than our interpolated value.  For P9155 Eq. 1 predicts $Y$=
12.285, while Hillenbrand et al. (2002) give $Y$ = 12.363 $\pm$
0.05, which is 0.08 mag fainter or 1.8-$\sigma$ different.

\parindent = 9 mm

We derived the $Y$-band magnitudes of the field stars listed in
Table 1 using observations taken on 6 nights (1999ee)  and 3
nights (2000bh), respectively, with 3 to 5 Persson stars per night,
whose $Y$-band magnitudes we derived using Eq. 1, rather than
relying on only one Persson star per night whose $Y$ magnitude
happens to be given by Hillenbrand et al.  Our $Y$-band photometry
of SNe may be subject to systematic errors, on the order of 0.03
mag, perhaps larger.  In any case, for the first time we present
$Y$-band light curves of Type Ia SNe.

The field star magnitudes given by us in Table 1 were checked against the
corresponding 2MASS values using {\sc simbad}. Limiting ourselves to field
stars brighter than magnitude 15, beyond which 2MASS data have large
uncertainties, we find that our $J_s$-band magnitudes are systematically
fainter than 2MASS by 0.014 mag, with a 1-$\sigma$ scatter of $\pm$ 0.026
mag.  For $H$ our field star magnitudes are systematically fainter by
0.028 mag, with $\sigma_H$ = $\pm$ 0.046 mag. For $K_s$ our field star
magnitudes are 0.030 fainter than 2MASS, with $\sigma_K$ = $\pm$ 0.059
mag.  Thus, there are no serious systematic differences between our field
star photometry and 2MASS values.

\subsection{SN 1999ee}

The Type Ia SN 1999ee was discovered on 1999 October 7.15 UT (JD
2,451,458.65) by Wischnjewsky as part of the SOIRS project and reported by
Maza et al. (1999).  It was located 10 arcsec east
and 10 arcsec south of the core of the SA(rs)bc-type galaxy IC
5179.  CCD photometry reported by Stritzinger et al. (2002) was
begun the following day; these authors give JD 2,451,469.1 $\pm$
0.5 as the time of $B$-band maximum.  Our infrared observations
were begun on JD 2,451,461.6, some 7.5 days before T($B_{max}$).
From optical data only, using analysis similar to that of Phillips
et al (1999), Stritzinger et al. (2002) derived A$_V$ = 0.94 $\pm$
0.16 for SN~1999ee. Extensive spectroscopy of this SN is discussed 
by Hamuy et al. (2002).

While observations of SN~1999ee were ongoing, another SN exploded in
the same galaxy, the Type Ib/c SN~1999ex.  It was discovered by Martin et al.
(1999) on 1999 November 9.51 UT.  Stritzinger et al. (2002) were able to 
detect SN~1999ex in pre-discovery images as early as 30 October
UT.  We will present infrared photometry of this SN in another paper.

In Table 1 we give infrared photometry of some of the field stars
near IC 5179.  The mean values are based on data taken on six
photometric nights.  We retain the numbering scheme of Stritzinger
et al. (2002). Star 3 is clearly the best field star for
calibration of the SN light curves, as it is close enough to the SN
to be on the detector at all times that the SN was.  Star 1 of
Stritzinger et al. was saturated on many of our LCO 1-m frames, so we have
not used it as a secondary standard.  Star 9 is quite faint and
star 6 is at the very edge of the IR mosaics, in which case it is
only useful on photometric nights.  Star 17 exhibited evidence of
some level of variability.  Because we decided to tie the
photometry of SN~1999ee to star 3 only, we have added
in quadrature 0.03, 0.02, 0.02, and 0.02 mag, respectively, to
uncertainties of the resulting $Y$, $J_s$, $H$, and $K_s$
photometry.

In Table 2 we give IR photometry of SN~1999ee from LCO, and in Table 3 we give
the YALO data along with corrections to the photometric system of Persson et al.
(1998).  For the LCO and YALO imagery we derived the photometry from {\sc psf}
magnitudes, using photometric templates obtained on 2000 November 15-17 UT, more
than 394 days after the time of $B$-band maximum.  We assume SN~1999ee had faded
away sufficiently after 13 months that there are no serious systematic errors
resulting from the image subtraction. Given that SN~1999ee had faded by 4.9 mag
in the Y-band by 2000 April 28 and that it was not detectable in J$_s$, H, or
K$_s$ in late April of 2000, the mid-November 2000 templates should be
acceptable.

The YALO $J$ and $H$ photometry has been corrected to the system of
$J_s$ and $H$ magnitudes of Persson et al. (1998) using the method
described by Stritzinger et al. (2002) and Krisciunas et al. (2003).
We used actual spectra of SN~1999ee described by Hamuy et al. (2002).
Fig. 6 of Krisciunas et al. (2003) shows the derived filter corrections,
which are added to the YALO IR photometry to place it on the system
of Persson et al. (1998).  These corrections amount to as much as 0.13 mag,
and from 10 to 30 d after T($B_{max}$) make the YALO $J$ and $H$ magnitudes
{\em brighter}.

The derivation of the filter corrections involves the construction of 
{\em effective} filter transmission profiles, each of which includes 
(as a function of wavelength) the atmospheric transmission, the nominal
filter transmission, aluminum reflections for the telescope
and within the instrument, window transmission for the instrument,
dichroic transmission (if the instrument contains one), and the
quantum efficiency of the detector.  The effective filter transmission
is meant to mimic the natural system of the observations.  This natural 
system is transformed to the standard system of Persson et al. (1998)
using color terms measured elsewhere.

As a consistency check, the $J$-band
color term from synthetic photometry of Vega, Sirius, and the Sun 
is $-$0.037 (scaling $J-H$), which compares well to the actual color
term from observations of Persson et al. (1998) standards of $-$0.043.
For the $H$-band the color term from synthetic photometry is +0.020,
also scaling $J-H$, which compares well with the actual value from 
observations of stars of +0.015.  These tests show that our model
bandpasses provide a reasonable match to the actual YALO bandpasses.

%

In Fig. \ref{99ee_yjhk} we show near-infrared light curves of SN~1999ee.
In all infrared bands the time of maximum was about 4 days prior 
to T($B_{max}$).

The reader will note that the corrected YALO photometry still does
not quite agree with the LCO photometry.  From about 15 to 20 days
after T($B_{max}$) the corrected YALO values are still fainter
than the LCO values. We discovered that we could reconcile the two
sets of data if we shifted the YALO $J$ filter 225
\AA\ to longer wavelengths and shifted the YALO $H$ filter 200
\AA\ to shorter wavelengths.  The synthetic color term for $H$
(+0.035) would still be in reasonable agreement with the actual
value (+0.015), given uncertainties of 0.01 to 0.02, but for the
$J$-band there would be considerable disagreement ($-$0.112 vs.
$-$0.043).  This shows that a simple wavelength shift does not
account for the differences in the photometry and that the shape
of the model bandpasses must be different than the actual ones,
or that the spectrophotometry of SN~1999ee is in error.
The conservative user of our photometry should
give greater weight to the LCO photometry, since it was primarily
taken with the very telescope used to set up the photometric
system of Persson et al. (1998).

\subsection{SN 2000bh}

SN 2000bh was discovered on 2000 April 5.23 UT (JD 2,451,639.73) by
Antezana as part of the SOIRS project.  The discoverey was reported by
Maza et al. (2000a).  SN~2000bh was located 8 arcsec west and 11 arcsec south
of the nucleus of ESO 573-14.  A spectrum taken on April 7 UT by L. Ho revealed
it to be a Type Ia SN near maximum light.

Fig. \ref{sn2000bh} shows the field of SN~2000bh and the nearby
field stars.

Infrared photometry of some of the fields stars near SN~2000bh is to be
found in Table 1.  Optical photometry of the nearby field stars is
given in Table 4.  Optical and infrared photometry of SN~2000bh is
given in Tables 5 and 6, respectively.  Because stars 2 and 3
were at the edges of the IR mosaics, we have tied the IR photometry
to stars 1 and 11 only.

In Fig. \ref{00bh_opt_ir} we show the optical and IR light curves of SN~2000bh. We
have added $BVI$ fits derived using the $\Delta$m$_{15}$ method (Phillips et al.
1999).

We find from analysis similar to Phillips et al. (1999) that 
$\Delta$m$_{15}$($B$) = 1.16 $\pm$ 0.10, T($B_{max}$) = JD 2,451,635.29
$\pm$ 0.28, E($B-V$)$_{host}$ = 0.02 $\pm$ 0.03, and m$-$M = 35.01 $\pm$ 0.08
on the Freedman et al. (2001) distance scale.

In Fig. \ref{00bh_opt_ir} we also show near infrared light curves of SN~2000bh.
If the behavior of SN~2000bh were like other Type Ia SNe, its IR maxima
occurred about JD 2,451,632, some 8 d before the start of our observations.
We note the extremely deep dip in the $J_s$-band light curve.  We also note that
the $H$-band secondary maximum must have been comparable in brightness to the first 
$H$-band maximum; this is similar to the behavior of SN~2001el (Krisciunas
et al. 2003).  


\subsection{SN 2000ca}

SN 2000ca was discovered as part of the SOIRS project by Antezana on 2000 
April 28.18 UT (JD 2,451,662.68) roughly 1 arcsec east and 5 arcsec north of 
the nucleus of ESO 383-32.  The discovery was reported by Maza et al. (2000b). 
A spectrum taken by Aldering \& Conley (2000) on April 29.3 UT
revealed SN~ 2000ca to be a Type Ia SN near maximum light.  From the ratio of Si II
lines at 580 and 610 nm, Aldering \& Conley suggested that this 
SN was slightly hotter and more luminous than normal; also, that the
equivalent width (0.02 nm) due to Na D at the redshift of the host galaxy
indicated modest extinction by the host galaxy.

Fig. \ref{sn2000ca} shows the field of SN~2000ca and the nearby field stars.

Infrared photometry of some of the fields stars near SN~2000ca is given
in Table 1.  Optical photometry of the nearby field stars is given
in Table 4.  Optical and infrared photometry of SN~2000ca is
to be found in Tables 7 and 8, respectively.  For the most part we
found that photometry based on image subtraction was very close (0.01 mag) to
photometry based on {\sc psf} magnitudes without image subtraction, but
to be on the safe side we preferred to derive our results from image 
subtractions.

The optical photometry was derived using image subtraction templates
obtained with the CTIO 1.5-m telescope on 2003 February 1 UT.  Infrared
photometry based on images taken with the LCO 1-m telescope was derived
using subtraction templates obtained on 2003 March 9 UT.  Infrared
photometry based on images taken with the LCO 2.5-m telescope was obtained
from {\sc psf} magnitudes without the use of $J_s$- and $H$-band subtraction
templates. 

In Fig. \ref{00ca_opt_ir} we show $UBVRIz$ optical and $J_sH$ light curves of
SN~2000ca.  We find from analysis similar to Phillips et al. (1999) that 
$\Delta$m$_{15}$($B$) = 0.98 $\pm$ 0.05, T($B_{max}$) = JD 2,451,666.22
$\pm$ 0.54, E($B-V$)$_{host}$ = 0.00 $\pm$ 0.03, and m$-$M = 34.91 $\pm$ 0.08
on the Freedman et al. (2001) distance scale.

\subsection{SN 2001ba}

SN 2001ba was discovered by Chassagne (2001) from images taken on 
2001 April 27.8 and 28.7 UT (Julian Dates 2,452,027.3 and 2,452,028.2).
It was located 19 arcsec east and 22 arcsec
south of the nucleus of MCG-05-28-1.  A spectrum by Nugent \& Wang
(2000) obtained on April 30 UT revealed it to be a Type Ia SN near
maximum light.

Fig. \ref{sn2001ba} shows SN~2001ba and the nearby field stars.
Optical and infrared photometry of
the field stars is given in Tables 4 and 1, respectively.  The field
stars were calibrated from observations with the LCO 1-m telescope.
All of the optical data of SN~2001ba itself were obtained with the YALO
1-m telescope.  From 20 nights of YALO data we determined
mean color coefficients, using our LCO calibration of the field stars 
near SN~2001ba as local ``standards''.  We then used these mean color terms
to determine photometric zero points for the transformation of the
instrumental {\sc psf} magnitudes of the SN to standardized values.  This resulted
in the smoothest possible light curves.

Optical photometry from the YALO 1-m telescope is given in Table 9. 
IR photometry from the LCO 1-m and 2.5-m telescopes is given in Table 10.
The optical and infrared light curves of SN~2001ba are shown in Fig.
\ref{01ba_opt_ir}.

We find from analysis similar to Phillips et al. (1999) that 
$\Delta$m$_{15}$($B$) = 0.97 $\pm$ 0.05, T($B_{max}$) = JD 2,452,034.48
$\pm$ 0.68, E($B-V$)$_{host}$ = 0.04 $\pm$ 0.03, and m$-$M = 35.55 $\pm$ 0.07 
on the Freedman et al. (2001) distance scale.

We have not devised filter corrections to place the YALO optical photometry of
SN~2001ba on the system of Bessel (1990). As explained in our paper on SN~2001el
(Krisciunas et al. 2003), the $B-V$ colors in the tail of the color curve will
likely be ``too red'' by $\approx$ 0.1 mag as a result.  This systematic
difference was taken into account when calculating E($B-V$)$_{tail}$ as part
of the $\Delta$m$_{15}$ analysis.

\section{Discussion}

\subsection {Determining maximum magnitudes in the infrared}

It is well known that the optical light curves of Type Ia SNe show various
patterns.  For example, if we order the light curves of different Type Ia SNe
by the decline rate parameter $\Delta$m$_{15}$($B$), the MLCS luminosity
parameter $\Delta$ or the Perlmutter et al. (1997) stretch factor, we see that
the secondary hump in the $I$-band decreases in strength and occurs closer to
the time of $B$-band maximum as we move from slower to faster decliners.  
Narrower light curves in $B$ and $V$ have weaker $I$-band secondary humps
(Hamuy et al. 1996b).  
For the fastest decliners such as SNe~1991bg and 1999by there is no secondary
hump at all (Garnavich et al. 2001, and references therein).  For the $R$-band
one typically sees a ``shoulder'' in the light curves of the slow decliners,
but no shoulder for the fast decliners.

Elias et al. (1985), Meikle (2000), Krisciunas et al. (2000), Verdugo et al.
(2002), and Phillips et al. (2003) have sought templates for the $JHK$ light
curves, with mixed results.  Candia et al. (2003) also give some $JHK$ light
curves, ordered by $\Delta$m$_{15}$($B$).  Along with the data presented in
this paper, we can state that the morphology of the infrared light curves does
not change with the same obvious pattern as the $R$-band and $I$-band light
curves. Roughly speaking, the {\em time} of the secondary $J$-band hump
moves to earlier phases for faster decliners. The strength of the dip in the
$J$-band light curves roughly two weeks after T($B_{max}$) and the strength of
the secondary hump following are not a monotonic function of
$\Delta$m$_{15}$($B$).  Also, the flatness of the $H$-band light curves 10
days after T($B_{max}$) is not a simple function of $\Delta$m$_{15}$($B$).

Given the available data, we have not yet devised a method of fitting the
$JHK$ light curves from $t$ = $-$12 to +60 days with respect to the $B$-band
maximum. However, we show here that there {\em are} stretchable templates that
appear to fit the infrared light curves within $\approx$13 days of the times
of the infrared maxima.

First it is necessary to correct the infrared photometry from the
observer's frame to the SN rest frame.  See Appendix B for details
on the infrared K-corrections.  These corrections are given in
Table 11 and are to be {\em subtracted} from the data to correct them for
the Doppler shifting of the spectral energy distributions of the SNe with
respect to the filter profiles.

Jha (2002, Fig. 3.8) gives new relationships between $\Delta$m$_{15}$($B$) and
the stretch parameters for $V$ and $B$.  The reader will note that the stretch
parameter $s$ is defined as the factor by which one stretches a template to match
a $V$- or $B$-band light curve.  If we seek a {\em template},
we seek values of $s^{-1}$ to apply to {\em light curves}.

Using values of $\Delta$m$_{15}$($B$) from Phillips et al. (1999), Krisciunas
et al. (2001, 2003), Strolger et al. (2002), and this paper, we calculated the
inverse stretch factors ($s^{-1}$) from the $V$- and $B$-band regressions of
Jha (2002), averaging the two for each object.  The objects in question were
SNe~1980N, 1986G, 1998bu, 1999aw, 1999ee, 2000ca, 2001ba, and 2001el.  
All of these objects were observed overlapping the
times of IR maxima, or shortly thereafter.  This made it easy to estimate the
maximum infrared magnitudes.  

Our templates use ``stretched time'' as the time coordinate.  We 
transformed the dates of the observations to (Julian Date $-$ T($B_{max}$))
times $s^{-1}$ and took out the cosmological time dilation by dividing by (1 +
$z$), where $z$ is the redshift.

In Fig. \ref{j_light_curves} we show the superimposition of the K-corrected,
stretched $J$-band light curves with respect to maximum light.  While
the dispersion increases significantly after $t$ = +10 d, the stretched
light curves show very uniform behavior overlapping the time of maximum.

In Fig. \ref{jhk_maxima} we show the $JHK$ templates overlapping the
time of the infrared maxima.  The {\sc rms} residuals of the third order
polynomial fits are $\pm$ 0.062, 0.080, and 0.075 mag, respectively, for
$J$, $H$, and $K$.  In Table 12 we give polynomial coefficients so that
the reader may use the fits shown in Fig. \ref{jhk_maxima} for the
extrapolation to maximum light for other supernovae.

The reader might naturally wonder: does the stretch method work at all for the
$I$-band? In Fig. \ref{i_stretch} we show the data for SNe~1998bu, 1999aw,
1999ee, 2000ca, 2001ba, and 2001el treated in the same manner as our $JHK$
data.  Clearly, the stretch method does work at the time of the $I$-band
maximum.  The {\sc rms} scatter in the plot is only $\pm$ 0.045 mag.

We are particularly encouraged by results shown in Fig. \ref{jhk_maxima},
because they give us a tool with which to estimate the maximum infrared
magnitudes of other Type Ia SNe which have many fewer data points in the window
of stretched time from $-$12 to +10 days.  In particular, Krisciunas, Phillips,
\& Suntzeff (2003) use the extinction-corrected maximum magnitudes of the
template objects listed above, along with data of several other Type Ia SNe to
produce $JHK$ Hubble diagrams.  They further show that there are no obvious
decline rate relations if we plot the resultant absolute magnitudes of the SNe
vs. $\Delta$m$_{15}$($B$).  Thus, Type Ia SNe in the infrared may be {\em
standard candles when at maximum brightness}.

\subsection {Color curves and extinction}

Until recently, extinction corrections for SNe have primarily been
calculated from a $B-V$ color excess and an assumed value of R$_V$, where
A$_V$ = R$_V \; \times$ E($B-V$).  Standard Galactic dust gives R$_V$ =
3.08 $\pm$ 0.15 (Sneden et al. 1978).  But there is no universal value of
R$_V$.  For example, in the case of SN~1999cl Krisciunas et al. (2000)
found that R$_V$ = 1.8.  Similarly, Hough et al. (1987) found that R$_V =
2.4 \pm 0.13$ may be appropriate for the dust in NGC 5128, the host of
SN~1986G.  Now that supernovae are regularly observed at optical and
infrared wavelengths, we can use a variety of color indices to calculate
A$_V$, perhaps with fewer worries about systematic errors.

In our first paper on optical and IR photometry of Type Ia SNe (Krisciunas
et al. 2000) we constructed unreddened loci of objects with mid-range
decline rates.  These loci were based on very few data points {\em per object}.
However, the well sampled photometry of SN~2001el and color curves based
on a generic H\"{o}flich model (Krisciunas et al. 2003, \S 3.4) showed that our 
original $V-H$ and $V-K$ unreddened loci contain no serious systematic errors.
This is to say that our empirical color curves and the synthetic color curves
based on a SN model are {\em consistent} with each other.  As stated by Krisciunas
et al. (2003), the agreement is better than the accuracy of the model.
The bottom line is that we believe the unreddened loci derived by
Krisciunas et al. (2000) allow us to determine A$_V$ to better than 0.1 mag
for Type Ia SNe of mid-range decline rates.

Over the past four years we have obtained enough well sampled infrared light
curves of Type Ia SNe that we may now construct the unreddened loci for
Type Ia SNe which are slow decliners.  Also, we can address the question
of the ``uniformity'' of the color curves for the slow decliners.  Candia
et al. (2003) gave preliminary $V-J$ and $V-H$ loci for the slow decliners.
What follows supercedes that work.

Krisciunas et al. (2000, 2001) previously considered objects of relatively low
redshift $z \lesssim$ 0.01).  Since we are now investigating objects up to $z$
= 0.038, it is necessary first of all to apply K-corrections to the $VJHK$
data.  For the $V$-band data we interpolated the values in Table 4 of Hamuy et
al. (1993). For $JHK$ we used the values presented here in Table 11.


If we adopt the values of A$_{\lambda}$ / A$_V$ =
0.282, 0.190, and 0.114 for the $J$-, $H$-, and $K$-bands, respectively,
given by Cardelli, Clayton, \& Mathis (1989; Table 3, column 5)
and assign an uncertainty of 20 percent to each ratio (to account for
some range of dust properties in {\em other} galaxies), it follows that:

\begin{equation}
A_V \; = \; (1.393 \pm 0.110) \; E(V-J) \; .
\end{equation}

\begin{equation}
A_V \; = (1.235 \pm 0.058) \; E(V-H) \; .
\end{equation}

\begin{equation}
A_V \; = (1.129 \pm 0.029) \; E(V-K) \; .  
\end{equation}



In Fig. \ref{vjhk} we show dereddened $V \; minus$ near infrared colors
of 1999aa, 1999aw, 1999ee, 1999gp, 2000ca, and 2001ba, whose decline
rates ranged from $\Delta$m$_{15}$($B$) = 0.81 to 1.00.\footnote[18]{In
the case of SN~1999aa we use the updated value (J. L. Prieto, private
communication) of $\Delta$m$_{15}$($B$) = 0.81 $\pm$ 0.04 based on the
data of Krisciunas et al. (2000) and Jha (2002).} Our adopted values of
A$_V$ were derived from the optical light curves using the method of
Phillips et al. (1999), and our values of $V \; minus \; $ IR color
excesses were derived using Eqs. 2 to 4. The data shown in Fig.
\ref{vjhk} are corrected for K-terms, extinction and time dilation.

In Table 13 we give least-squares regressions to subsets of the data shown
in Fig. \ref{vjhk}.  The reader should pay particular attention to the
range of time over which we feel the fits are valid.

From $-8 \lesssim t \lesssim +9.5$ d the dereddened $V-J$ colors get
monotonically bluer.  The scatter about the line is only $\pm$ 0.067 mag.  
Thus, from the available data, $V-J$ colors of slow decliners may be
regarded as ``uniform'' over this range of time with respect to $B$-band
maximum. However, at $t$ = +9.5 d the dispersion of $V-J$ colors increases
significantly.

In Fig. \ref{vjhk} we show as dashed lines the fourth order fits to
the $V-H$ and $V-K$ colors of eight SNe of mid-range decline rates
considered by Krisciunas et al. (2000).  Coefficients to generate these
polynomials are given in Table 14.  These loci are consistent with
Peter H\"{o}flich's modeling presented in our paper on SN~2001el
(Krisciunas et al. 2003, \S 3.4). One obvious thing to note about the
slow decliners is that they have $V-H$ and $V-K$ colors which are
bluer than those of the mid-range decliners. At the time of $B$-band
maximum the slow decliners are 0.243 and 0.230 mag bluer, respectively,
in these two color indices.

In the case of $V-H$ colors, a fourth order fit to the dereddened colors
of SNe~1999aw, 1999ee, 1999gp, 2000ca, and 2001ba exhibits a scatter of
$\pm$ 0.062 mag with a reduced $\chi^2$ of 1.9 prior to $t$ = +8.5 d.
However, the scatter from $+8.5 \leq t \leq +27$ d is $\pm$ 0.15 mag
with $\chi^2_{\nu}$ = 12.  Thus, $V-H$ colors of slow decliners may be
considered ``uniform'' within 8 or 9 days of T($B_{max}$), but {\em not}
afterwards.

Our dereddened $V-K$ colors of {\em slow} decliners do not exhibit any
epoch over which the {\sc rms} scatter is less than $\pm$ 0.1 mag.  
The $V-K$ colors of SN~1999ee, for example, are roughly 0.3 mag redder
than those of SN~2001ba, and these two objects have quite similar
$B$-band decline rates.  While the formal scatter of the fourth order
fit to the data of SNe 1999aa, 1999aw, 1999ee, 1999gp, and 2001ba is
$\pm$ 0.138 mag, we do not think that the available data allow us to
assert that the $V-K$ colors of the slow decliners are particularly
uniform.  The reader should notice from Table 11 that the
K-corrections are quite large for the LCO $K_s$-band and a strong
function of the redshift. Some of the spread of the $V-K$ colors shown
in the bottom panel of Fig. \ref{vjhk} may be due to differences in
the $K$-band spectra of the SNe, which is to say that our
K-corrections, based on SN~1999ee but applied to different objects,
may add systematic error to the results.  See Figs. 7 and 8 of Hamuy
et al. (2002) for a comparison of spectra of different SNe.
{\em Some} of the spread of the photometry must be due to the
uncertainties of the reddening coefficients in Eqs. 2 to 4. 

We wish to consider if there is evidence of a {\em continuous} change of $V
\; minus \; $ IR colors at some epoch.  One can often determine a maximum
magnitude more accurately than the {\em time} of maximum. As a result, we
show in Fig. \ref{vjhkmax} the dereddened colors $V_{max}-J_{max}$,
$V_{max}-H_{max}$, and $V_{max}-K_{max}$.\footnote[19] {The reader will note
that because the $V-$band maximum of a Type Ia SN occurs on average about
five to six days after the IR maximum, the colors plotted in Fig.
\ref{vjhkmax} are unphysical in the sense that they do not correspond to
observations that can be made at a single moment in time.} The objects
considered are SNe 1980N (Hamuy et al. 1991), 1981B (Buta \& Turner 1983),
1998bu (Suntzeff et al. 1999, Jha et al. 1999), 
Phillips et al. 2003), 
1999ee (Stritzinger et al. 2002), 2000bh, 2000bk (Krisciunas et al. 2001),
2000ca, 2001ba, and 2001el (Krisciunas et al. 2003).  We also consider
SNe~1994D (Richmond et al. 1995), 1999gp and 2000ce (Krisciunas et al. 2001),
which had very few data points in the [$-$12,+10] day window. The IR data can
be found in the papers cited, in Meikle (2000), and in the tables of this
paper.

Since relatively few SNe are discovered early enough to measure their
infrared maxima {\em directly}, we plot in Fig. \ref{vjh6} the {\em
interpolated} $V-J$ and $V-H$ colors 6 days after $B$-band maximum.  
The regression lines weighted by the errors of the points are:

\begin{equation}
(V-J)_{t=+6} \; = \; (-1.837 \pm 0.148) + (0.570 \pm 0.138) \; \times \; 
\Delta m_{15}(B) \; . 
\end{equation}

\begin{equation}
(V-H)_{t=+6} \; = \; (-1.836 \pm 0.169) + (0.650 \pm 0.157) \; \times \; 
\Delta m_{15}(B) \; . 
\end{equation}

\parindent = 0 mm

The {\sc rms} residual of the $V-J$ fit is $\pm$ 0.125 mag, with a
reduced $\chi^2$ value of 1.22, while for $V-H$, $\sigma$({\sc rms}) =
$\pm$ 0.146 mag and $\chi^2_{\nu}$ = 1.32.  The scatter corresponds to
to an uncertainty in A$_V$ of $\pm$ 0.18 mag, or $\pm$ 0.06 mag
in E($B-V$). This is comparable to the advertised accuracy of determining
extinction and reddening with the $\Delta$m$_{15}$ method.

\parindent = 9 mm

As time goes on our growing database will allow us to make plots analogous to
Figs. \ref{vjhkmax} and \ref{vjh6} which will include {\em only} those
objects with small host extinction and whose light curves were well sampled
at maximum.  These two preliminary figures can be interpreted to mean that
there {\em is} evidence for a continuous change of $V-J$ and $V-H$ color as a
function of decline rate. $V-K$ data may eventually show a similar
trend.

Regarding $V$ {\em minus} IR loci for the mid-range decliners and the slow
decliners, the presently available data can be looked at two ways.  It is
equally valid to average the data of several objects over some range of
decline rates, {\em or} to assert qualitatively that there is a continuous
change of color as a function of decline rate at some epoch earlier than $t$
= +9 d. The idea of continuous color variation as a function of decline rate
is more pleasing aesthetically, because one would think that cosmic
explosions exhibit a continuous range of energies, temperatures, and
opacities.

\parindent = 9 mm




\section{Conclusions}

In this paper we presented well sampled light curves for four Type Ia
SNe. This includes the first-ever SN data published at 1.03 $\mu$m,
though the $Y$-band data lack a firm basis in calibration.  Given that
the maxima in the infrared light curves occur typically 3 to 4 days
before the time of $B$-band maximum, it is a considerable challenge to
obtain light curves that cover the IR maxima.  We achieved this in the
cases of SNe~1999ee, 2000ca, and 2001ba.

We have now observed a sufficient number of Type Ia SNe near the time of
maximum that we can investigate the question of uniform templates for the
infrared.  If we stretch the light curves in the time domain according to
stretch factors based on the $B$- and $V$-band relationships of Jha (2002,
Fig. 3.8), we obtain reasonably uniform templates from $-$12 to +10 days
after maximum (in ``stretched days'').  The {\sc rms} uncertainties of
the fits are $\pm$~0.062, 0.080, and 0.075 mag in $J$, $H$, and $K$,
respectively.  This allows us to determine the maximum magnitudes
of other Type Ia SNe as long as we have some data in the $-$12 to
+10~d window, and the light curves are not obviously ``abnormal''
(e.g. like SN~2000cx).

Primarily on the basis of observations carried out at CTIO and Las
Campanas since the beginning of 1999, we can now draw the $V \; minus$
IR color curves for Type Ia SNe which are mid-range decliners and slow
decliners.  We can utilize the present observational data, backed up
by modeling for the mid-range decliners (Krisciunas et al. 2003, \S
3.4), to determine the extinction suffered by these objects.  This is
true even for highly reddened objects that occurred in galaxies with
different dust properties than we measure for the dust in our Galaxy.

The motivation behind using IR data is that extinction is considerably
lower than in the optical, and if we know the intrinsic $V \; minus$ IR
colors of Type Ia SNe, then we can determine A$_V$ by slightly scaling the
$V \; minus$ IR color excesses.  This should give us more accurate
extinction corrections compared to deriving $B-V$ color excesses and
scaling them by an assumed value of R$_V$.  However, while we can
enumerate unreddened loci for Type Ia SNe of mid-range and slow decline
rates, for the slow decliners the ``uniformity'' of $V-J$ and $V-H$ color
is limited to within nine days of the time of $B$-band maximum, just
about what we found for the construction of our $JHK$ templates.

The available data can also be used as evidence that there is a {\em
continuous} change of $V \; minus \; $ IR color at $t$ = +6 d vs. the
decline rate.  The $V_{max}-J_{max}$ and $V_{max}-H_{max}$ colors are
further evidence that Type Ia SNe are progressively redder as a function
of decline rate.  Given that we already have evidence for a spectroscopic
sequence as function of temperature (Nugent et al. 1995), the further
calibration of the optical-IR color relationships will be relevant to SN
modeling.

As large scale surveys of the sky are planned and implemented, we
can and should make use of the light curve and color templates
presented here to derive the maximum magnitudes of Type Ia SNe
and correct the data for the effects of dust along the line of
sight.  Since Type Ia SNe are the most important distance indicator
at redshifts $z \; >$ 0.01, we may further solidify the foundation
of the cosmological distance ladder with these useful cosmic beacons.



\acknowledgements  

We made extensive use of the NASA/IPAC Extragalactic Database (NED).  We
also made use of the {\sc simbad} database, operated at CDS, Strasbourg,
France.  We thank the Space Telescope Science Institute for the following
support: HST GO-07505.02A; HST GO-08177.06 (the High-Z Supernova Team
survey); HST GO-08641.07A was provided by NASA through a grant from the
Space Telescope Science Institute, which is operated by the Association of
Universities for Research in Astronomy, Inc., under NASA contract
NAS5-26555. We thank STScI for the salary support for PC from grants GO-09114 and
GO-09421.  We thank Arlo Landolt for many useful discussions. 
Our late colleague Robert Schommer was a strong
supporter of the YALO 1-m telescope, with which we have obtained a large
amount of supernova data.  KK thanks LCO and NOAO for funding part of his
postdoctoral position.

\appendix

\section{Calibration of $z$-band Photometry}

Hamuy (2001, Appendix B) gives synthetic $z$-band magnitudes of 20 of the
spectrophotometric standards of Stone \& Baldwin (1983).
Using the $VRI$ photometry of 17 of these stars given by
Landolt (1992b), we obtain the following regression:

\begin {equation}
(R-z) \; = \; (-0.0416 \pm 0.0069) \; + (0.7627 \pm 0.0140) \; 
(V-I) \; , 
\end {equation}

\parindent = 0 mm

with an {\sc rms} residual of $\pm$ 0.020 mag.  The range of
$V-I$ color of this sample is $-0.266$ to +0.769.

\parindent = 9 mm

From 10 to 12 November 2001 and 30 January to 1 February 2003 we
imaged a number of Landolt (1992a) fields in $VRIz$ using the
CTIO 1.5-m telescope and also
imaged some of the Stone-Baldwin standards.  In Fig. \ref{vi_rz} 
we show the $R-z$ vs. $V-I$ colors. 

The solid line in Fig. \ref{vi_rz} is that of Eq. A1 above.
Over a similar color range of the Stone-Baldwin standards, the
{\sc rms} scatter of $R-z$ colors of the Landolt (1992a) standards 
is $\pm$ 0.039 mag.  Some of this larger scatter is due to three stars 
whose points lie noticeably above the line: Rubin 149C, SA95-96, and Rubin 152E.
We offer no explanation for these deviations, but note that these
three stars have $B-V$ colors like those of stars of spectral type A.

Somewhere around $V-I$ = 1.00 there is a change of slope.  From the 
points with $V-I > $ 0.97 we find:

\begin {equation}
(R-z) \; = \; (-0.0196 \pm 0.0289) \; + (0.7191 \pm 0.0217) \;
(V-I) \; ,
\end {equation}

\parindent = 0 mm

with an {\sc rms} residual of $\pm$ 0.030 mag. 

\parindent = 9 mm

Naturally, if one wishes to calibrate $z$-band photometry, it is best
to use the Stone-Baldwin standards, but with the three exceptions noted
above, there is a rather tight relationship between the $R-z$ and $V-I$
colors of Landolt (1992a) standards.

\section{Infrared K-corrections}

The K-corrections for the near-infrared bands $J_sHK_s$ have been
calculated in the manner described in Hamuy et al. (1993). The
K-correction is given with the usual sign convention: $m_i(z=0) =
m_i(z) - K_i(z)$ where $m_i(z)$ is the magnitude as observed at the
telescope and $m_i(z=0)$ is the corrected magnitude as if it were
observed in the frame of the supernova. Optical K-corrections are
typically positive and an increasing function in $z$ because the
optical flux of most sources increases towards the red. The
K-correction vanishes for $F_\lambda \propto \lambda^{-1}$ independent
of the filter. Since the near-IR flux tends to decrease as function of
wavelength, most of the near-IR K-corrections are negative. The
K-corrections are not as monotonic in $z$ as in the optical because
large nebular emission features dominate the continuum flux.

These K-corrections are based on 11 spectra of a single supernova,
SN~1999ee (Hamuy et al. 2002), spanning days $-$9 to +41 from
T($B_{max}$). The spectra were corrected to zero heliocentric velocity
from the observed velocity of $v$ = 3498 km s$^{-1}$. Each spectrum was
inspected by eye. For the regions between the $J$/$H$ and $H$/$K$ bands
where there is significant water absorption, a simple linear
interpolation was made. In some cases, we extended the spectrum on a
given date by logrithmic interpolation between the spectra bracketing the
given spectrum in time. This was generally done to extend a spectrum
slightly to the red in order for the spectrum to overlap ``unredshifted''
filter curve.

The filter transmission functions were calculated from the filter
curves given in Persson et al. (1998) for the $J_s$, $H$, and $K_s$
filters. These filter functions were mulitplied by a standard atmosphere
at airmass = 1 (to introduce the saturated telluric absorption), two
aluminum reflections (to mimic the telescope mirrors), three aluminum
reflections (to mimic the internal reflective optics), and a typical
quantum efficieny of a Rockwell HgCdTe detector.  The remaining
optical element $-$ the Dewar window $-$ was ignored.  The product of
these curves defines the natural system of the Persson et al.
standards.

Because these K-corrections are based on a single supernova, these
tables should be used with caution for supernovae that have
significantly different spectral features than SN~1999ee.

\clearpage

\begin{deluxetable}{ccccccc}
\tablecolumns{7}
\tablewidth{0pc} 
\tablecaption{Infrared Photometry of Field Stars}
\startdata 
Field   &    star  &  $Y$   &   $J_s$   &  $H$  &  $K_s$  & N$_{obs}$\tablenotemark{a}  \\ \tableline \tableline
1999ee  &    3    & 14.066 (0.008) & 13.646 (0.008) & 13.039 (0.010) & 12.756 (0.014) & 6 6 6 6 \\
\ldots  &    6    & 15.400 (0.006) & 15.025 (0.010) & 14.410 (0.023) & 14.204 (0.028) & 4 6 6 6 \\
\ldots  &    9    & 17.686 (0.049) & 16.954 (0.024) & 16.359 (0.033) & 16.220 (0.045) & 5 5 6 5 \\
\ldots  &    17   & 14.531 (0.016) & 14.080 (0.022) & 13.500 (0.008) & 13.231 (0.019) & 6 6 6 6 \\
2000bh  &    1    & 14.329 (0.007) & 14.070 (0.004) & 13.627 (0.005) & 13.533 (0.011) & 4 8 8 6 \\
\ldots  &    2    & 14.507 (0.016) & 14.227 (0.007) & 13.728 (0.009) & 13.702 (0.017) & 4 8 8 6 \\
\ldots  &    3    & 13.267 (0.022) & 12.796 (0.022) & 12.239 (0.011) & 12.003 (0.009) & 4 7 9 6 \\
\ldots  &    11   & 14.171 (0.015) & 13.952 (0.004) & 13.590 (0.005) & 13.495 (0.010) & 4 8 8 6 \\
2000ca  &    1    &     \nodata    & 14.135 (0.019) & 13.802 (0.033) &  \nodata       & 0 4 6 0 \\
\ldots  &    3    &     \nodata    & 16.231 (0.051) & 15.896 (0.007) &  \nodata       & 0 4 6 0 \\
2001ba  &    1    &     \nodata    & 14.679 (0.007) & 14.339 (0.010) & 14.323 (0.016) & 0 4 4 3 \\
\ldots  &    2    &     \nodata    & 16.363 (0.017) & 16.065 (0.037) & 16.060 (0.047) & 0 4 4 3 \\
\ldots  &    20   &     \nodata    & 15.862 (0.005) & 15.460 (0.037) & 15.455 (0.025) & 0 4 4 3 \\
\enddata
\tablenotetext{a} {Number of observations that were used to determine the
 weighted means, for filters $Y$, $J_s$, $H$, and $K_s$, respectively.}
\end{deluxetable}

\begin{deluxetable}{cccccc}
\tablecolumns{7}
\tablewidth{0pc}
\tablecaption{LCO Infrared Photometry of SN 1999ee}
\startdata
JD\tablenotemark{a}  &  $Y$   &   $J_s$   &  $H$  &  $K_s$  & Telescope\tablenotemark{b}   \\ \tableline \tableline
461.59  &     \nodata    & 14.964 (0.022) & 15.117 (0.024) & 14.859  (0.041) & 1 \\
462.56	& 14.899 (0.032) & 14.804 (0.022) & 15.082 (0.024) & 14.808  (0.041) & 1 \\
463.55	& 14.788 (0.032) & 14.797 (0.022) & 15.016 (0.024) &      \nodata    & 1 \\
464.56	& 14.733 (0.032) & 14.669 (0.022) & 14.934 (0.024) & 14.667  (0.036) & 1 \\
465.52	& 14.811 (0.032) & 14.676 (0.022) & 14.993 (0.025) & 14.572  (0.034) & 1 \\
467.52	& 14.880 (0.032) & 14.817 (0.022) & 15.000 (0.026) & 14.626  (0.043) & 1 \\
468.53	& 14.928 (0.032) & 14.797 (0.022) & 15.121 (0.026) & 14.592  (0.042) & 1 \\
477.50  &     \nodata    &     \nodata    & 15.312 (0.026) &     \nodata     & 2 \\
478.51  &     \nodata    & 15.716 (0.024) &      \nodata   &     \nodata     & 2 \\
479.52  &     \nodata    &     \nodata    & 15.411 (0.028) &     \nodata     & 2 \\
481.48  &     \nodata    &      \nodata   & 15.391 (0.028) &     \nodata     & 2 \\
482.57	& 15.534 (0.033) & 16.175 (0.026) & 15.348 (0.030) & 14.748  (0.046) & 1 \\
483.56	& 15.588 (0.032) & 16.289 (0.030) & 15.349 (0.027) & 14.809  (0.060) & 1 \\
484.54	& 15.464 (0.032) & 16.355 (0.026) & 15.259 (0.027) & 14.918  (0.039) & 1 \\
485.50  &  \nodata       & 16.325 (0.033) & 15.179 (0.034) &    \nodata      & 1 \\
486.53	& 15.416 (0.032) & 16.314 (0.028) & 15.259 (0.026) & 14.853  (0.046) & 1 \\
487.55	& 15.419 (0.032) & 16.228 (0.026) & 15.128 (0.026) & 14.972  (0.046) & 1 \\
488.55	& 15.347 (0.032) & 16.285 (0.024) & 15.161 (0.025) & 14.961  (0.052) & 1 \\
489.56	& 15.279 (0.031) & 16.240 (0.026) & 15.176 (0.025) & 14.934  (0.050) & 1 \\
661.88\tablenotemark{c}  &  \nodata   & $>$ 19.00      & $>$ 18.64      & $>$ 17.00       & 1 \\
662.86  & 19.64 (+0.26,$-$0.21) &   \nodata &   \nodata    &  \nodata        & 1 \\
\enddata
\tablenotetext{a} {Julian Date $minus$ 2,451,000.}
\tablenotetext{b}{1 = LCO 1-m.  2 = LCO 2.5-m.} 
\tablenotetext{c}{The data of Julian Date 2,451,661 are 3-$\sigma$ upper limits.}
\end{deluxetable}

\begin{deluxetable}{ccccc}
\tablecolumns{7}
\tablewidth{0pc}
\tablecaption{YALO Infrared Photometry of SN 1999ee and Filter Corrections\tablenotemark{a}}
\startdata
JD\tablenotemark{b}  &   $J$   & $\Delta\; J$ &  $H$  &  $\Delta\; H$ \\ \tableline \tableline
461.60 & 15.006 (0.026) & \phs0.042  & 15.147 (0.034) & \phs0.003     \\
462.57 & 14.880 (0.026) & \phs0.044  & 15.071 (0.030) & \phs0.001     \\
464.62 & 14.743 (0.027) & \phs0.052  & 15.004 (0.031) & $-$0.003     \\
466.64 & 14.709 (0.026) & \phs0.058  & 15.032 (0.030) & $-$0.008     \\
468.63 & 14.800 (0.025) & \phs0.065  & 15.103 (0.028) & $-$0.012     \\

470.64 & 14.915 (0.025) & \phs0.068  & 15.186 (0.028) & $-$0.018    \\
472.65 & 15.072 (0.025) & \phs0.048  & 15.297 (0.031) & $-$0.033     \\
475.64 & 15.343 (0.029) & \phs0.019  & 15.354 (0.033) & $-$0.042     \\
478.66 & 15.745 (0.033) & $-$0.008  & 15.407 (0.037) & $-$0.044     \\
480.66 & 16.121 (0.030) & $-$0.025  & 15.520 (0.042) & $-$0.044     \\
\\
482.63 & 16.383 (0.033) & $-$0.041  & 15.500 (0.037) & $-$0.044     \\
484.57 & 16.487 (0.030) & $-$0.057  & 15.445 (0.033) & $-$0.045     \\
486.64 & 16.528 (0.040) & $-$0.070  & 15.356 (0.032) & $-$0.053     \\
488.64 & 16.521 (0.032) & $-$0.083  & 15.254 (0.034) & $-$0.060     \\
490.65 & 16.518 (0.031) & $-$0.096  & 15.230 (0.033) & $-$0.067     \\

492.63 & 16.411 (0.032) & $-$0.088  & 15.252 (0.038) & $-$0.071     \\
495.60 & 16.271 (0.031) & $-$0.055  & 15.189 (0.033) & $-$0.070     \\
497.55 & 16.209 (0.034) & $-$0.055  & 15.161 (0.034) & $-$0.058     \\
498.55 & 16.164 (0.034) & $-$0.064  & 15.195 (0.036) & $-$0.045     \\
499.57 & 16.047 (0.029) & $-$0.074  & 15.111 (0.031) & $-$0.030    \\
\\
500.55 & 15.978 (0.034) & $-$0.084  & 15.185 (0.030) & $-$0.019     \\
501.53 & 15.963 (0.030) & $-$0.087  & 15.161 (0.029) & $-$0.014    \\
503.53 & 15.878 (0.030) & $-$0.098  & 15.314 (0.036) & $-$0.003    \\
504.54 & 15.899 (0.029) & $-$0.103  & 15.268 (0.033) & \phs0.002     \\
505.53 & 15.922 (0.027) & $-$0.107  & 15.435 (0.038) & \phs0.009     \\

507.59 & 16.124 (0.028) & $-$0.117  & 15.513 (0.039) & \phs0.018     \\
508.55 & 16.216 (0.033) & $-$0.122  & 15.537 (0.045) & \phs0.023     \\
511.55 & 16.466 (0.039) & $-$0.131  & 15.792 (0.043) & \phs0.033     \\
513.60 & 16.552 (0.048) & $-$0.131  & 15.971 (0.055) & \phs0.033     \\
515.60 & 16.797 (0.038) & $-$0.131  & 15.975 (0.042) & \phs0.033     \\
\\
517.54 & 16.887 (0.040) & $-$0.131  & 16.067 (0.044) & \phs0.033     \\
519.54 & 17.129 (0.047) & $-$0.131  & 16.128 (0.054) & \phs0.033     \\
521.59 & 17.263 (0.063) & $-$0.131  & 16.208 (0.070) & \phs0.033     \\

\enddata
\tablenotetext{a} {The photometry corrections of columns 3 and 5, 
when {\em added} to the data given in columns 2 and 4, respectively,
place the YALO $JH$ data on the $J_sH$ system of Persson et al. (1998).
These corrections were derived from spectra of SN~1999ee (Hamuy et al. 2002)
and appropriate filter transmission functions.}
\tablenotetext{b}{Julian Date $minus$ 2,451,000.}
\end{deluxetable}

\begin{deluxetable}{cccccccc}
\tablecolumns{8}
\rotate
\tablewidth{0pc}
\tablecaption{Optical Photometry of Field Stars}
\startdata
Field   & star  &  $U$   &   $B$   &  $V$  &  $R$  & $I$ & $z$  \\ \tableline \tableline
%
%
2000bh & 1  & 16.946 (0.018) & 16.480 (0.006) & 15.617 (0.006) & 15.133 (0.006) & 14.692 (0.006) & 14.511 (0.023) \\ 
\ldots & 2  & 16.827 (0.017) & 16.609 (0.006) & 15.805 (0.006) & 15.330 (0.006) & 14.856 (0.006) & 14.662 (0.020) \\ 
\ldots & 3  & 20.063 (0.068) & 18.737 (0.020) & 17.024 (0.006) & 15.786 (0.007) & 14.222 (0.008) & 13.642 (0.062) \\ 
\ldots & 4  & \nodata        & 16.674 (0.006) & 15.799 (0.006) & 15.300 (0.006) & 14.801 (0.006) & 14.590 (0.027) \\
\ldots & 5  & 16.319 (0.027) & 16.373 (0.006) & 15.773 (0.006) & 15.408 (0.006) & 15.040 (0.006) & 14.898 (0.017) \\
\ldots & 10 & 18.743 (0.023) & 17.474 (0.008) & 16.225 (0.006) & 15.449 (0.006) & 14.749 (0.006) & 14.449 (0.033) \\
\ldots & 11 & 16.241 (0.031) & 16.021 (0.014) & 15.292 (0.006) & 14.885 (0.006) & 14.501 (0.009) & 14.353 (0.023) \\

2000ca & 1  & 15.576 (0.019) & 15.745 (0.012) & 15.225 (0.010) & 14.898 (0.013) & 14.556 (0.013) & 14.415 (0.018) \\
\ldots & 2  & 15.026 (0.025) & 14.990 (0.014) & 14.408 (0.012) & 14.061 (0.014) & 13.719 (0.013) & 13.593 (0.018) \\
\ldots & 3  & 18.386 (0.011) & 18.246 (0.007) & 17.554 (0.006) & 17.128 (0.020) & 16.754 (0.009) & 16.564 (0.028) \\
\ldots & 4  & 16.977 (0.011) & 16.922 (0.009) & 16.270 (0.008) & 15.892 (0.009) & 15.521 (0.011) & 15.353 (0.022) \\
\ldots & 5  & 16.775 (0.092) & 16.787 (0.010) & 16.167 (0.008) & 15.805 (0.009) & 15.450 (0.011) & 15.310 (0.022) \\
\ldots & 6  & 16.401 (0.020) & 15.946 (0.008) & 15.062 (0.006) & 14.567 (0.006) & 14.081 (0.008) & 13.863 (0.007) \\
\ldots & 7  & 17.935 (0.026) & 18.141 (0.007) & 17.721 (0.006) & 17.411 (0.012) & 17.089 (0.013) & 16.965 (0.053) \\

2001ba & 1  & \nodata & 16.517 (0.024) & 15.894 (0.017) & 15.505 (0.023) & 15.152 (0.024) & \nodata \\
\ldots & 2  & \nodata & 18.069 (0.033) & 17.499 (0.023) & 17.147 (0.026) & 16.789 (0.028) & \nodata \\
\ldots & 3  & \nodata & 15.967 (0.023) & 15.197 (0.016) & 14.768 (0.022) & 14.338 (0.023) & \nodata \\
\ldots & 4  & \nodata & 17.982 (0.031) & 16.990 (0.020) & 16.440 (0.023) & 15.897 (0.024) & \nodata \\
\ldots & 5  & \nodata & 17.908 (0.030) & 16.944 (0.019) & 16.384 (0.023) & 15.860 (0.024) & \nodata \\
\ldots & 6  & \nodata & 17.571 (0.028) & 16.768 (0.019) & 16.331 (0.023) & 15.862 (0.024) & \nodata \\
\ldots & 7  & \nodata & 16.623 (0.024) & 15.607 (0.016) & 15.042 (0.022) & 14.553 (0.023) & \nodata \\
\ldots & 8  & \nodata & 15.818 (0.023) & 15.131 (0.016) & 14.741 (0.022) & 14.337 (0.023) & \nodata \\
\ldots & 9  & \nodata & 15.938 (0.023) & 15.258 (0.016) & 14.899 (0.022) & 14.535 (0.023) & \nodata \\
\ldots & 10 & \nodata & 16.228 (0.024) & 15.709 (0.016) & 15.411 (0.022) & 15.095 (0.023) & \nodata \\
\ldots & 11 & \nodata & 16.940 (0.025) & 16.024 (0.017) & 15.568 (0.022) & 15.170 (0.023) & \nodata \\
\ldots & 14 & \nodata & 16.150 (0.024) & 15.254 (0.016) & 14.761 (0.022) & 14.276 (0.023) & \nodata \\
\ldots & 17 & \nodata & 14.767 (0.023) & 13.952 (0.016) & 13.558 (0.015) & 13.120 (0.030) & \nodata \\
\ldots & 19 & \nodata & 16.105 (0.024) & 15.412 (0.016) & 15.034 (0.022) & 14.670 (0.023) & \nodata \\
\ldots & 20 & \nodata & 17.866 (0.031) & 17.176 (0.021) & 16.781 (0.024) & 16.361 (0.025) & \nodata \\
\enddata
\end{deluxetable}

\begin{deluxetable}{cccccccc}
\tablecolumns{8}
\rotate
\tablewidth{0pc}
\tablecaption{Optical Photometry of SN 2000bh }
\startdata
JD\tablenotemark{a}  &  $U$  & $B$   &   $V$   &  $R$  &  $I$ &  $z$  & Telescope\tablenotemark{b}    \\ \tableline \tableline
641.72 &    \nodata     & 16.284 (0.018) & 16.103 (0.010) &   \nodata      & 16.513 (0.010) &    \nodata     & 2 \\
642.68 &    \nodata     & 16.405 (0.013) & 16.155 (0.015) &   \nodata      & 16.592 (0.011) &    \nodata     & 2 \\
643.62 & 16.632 (0.060) & 16.483 (0.012) & 16.192 (0.012) & 16.108 (0.012) & 16.633 (0.012) & 16.399 (0.020) & 1 \\
661.78 &    \nodata     & 18.335 (0.035) & 17.186 (0.012) & 16.810 (0.012) & 16.841 (0.012) &    \nodata     & 2 \\ 
675.57 &    \nodata     & 19.120 (0.013) & 18.009 (0.012) & 17.549 (0.012) & 17.366 (0.012) & 17.023 (0.020) & 2 \\
676.61 &    \nodata     & 19.183 (0.013) & 18.047 (0.012) & 17.609 (0.012) & 17.430 (0.012) &    \nodata     & 2 \\
677.62 &    \nodata     & 19.213 (0.026) & 18.108 (0.014) & 17.661 (0.018) & 17.476 (0.012) & 17.244 (0.020) & 2 \\
680.75 &    \nodata     & 19.310 (0.057) & 18.190 (0.016) &   \nodata      &                & 17.425 (0.020) & 2 \\
681.58 &    \nodata     & 19.296 (0.027) & 18.237 (0.013) & 17.822 (0.010) & 17.700 (0.014) & 17.412 (0.026) & 2 \\
682.52 &    \nodata     & 19.326 (0.031) & 18.234 (0.012) & 17.843 (0.012) & 17.757 (0.015) &    \nodata     & 2 \\
683.57 &    \nodata     & 19.330 (0.026) & 18.285 (0.035) & 17.874 (0.011) & 17.770 (0.015) & 17.608 (0.027) & 2 \\
684.57 &    \nodata     & 19.355 (0.023) & 18.304 (0.012) & 17.916 (0.010) & 17.824 (0.011) & 17.584 (0.021) & 2 \\
\enddata  
\tablenotetext{a} {Julian Date $minus$ 2,451,000.}
\tablenotetext{b} {1 = ESO NTT 3.6-m.  2 = CTIO 0.9-m.}
\end{deluxetable}

\begin{deluxetable}{cccccc}
\tablecolumns{6}
\tablewidth{0pc}
\tablecaption{Infrared Photometry of SN 2000bh }
\startdata
JD\tablenotemark{a}  &  $Y$   &   $J_s$   &  $H$  &  $K_s$  & Telescope\tablenotemark{b}   \\ \tableline \tableline
640.66 &    \nodata     & 16.799 (0.034) & 16.895 (0.030) & 16.636 (0.042) & 1 \\
641.61 &    \nodata     & 17.054 (0.029) & 17.011 (0.036) & 16.679 (0.046) & 1 \\
642.55 & 17.094 (0.033) & 17.080 (0.027) & 16.941 (0.030) & 16.831 (0.055) & 1 \\
643.60 &    \nodata     & 17.229 (0.027) & 16.993 (0.031) & 16.824 (0.059) & 1 \\
644.58 & 17.143 (0.032) & 17.442 (0.038) & 17.073 (0.030) & 16.855 (0.053) & 1 \\

650.57 &    \nodata     & 18.374 (0.044) & 17.245 (0.033) &    \nodata     & 1 \\
652.59 &    \nodata     & 18.315 (0.032) &     \nodata    &    \nodata     & 1 \\
653.61 & 17.133 (0.030) & 18.450 (0.040) & 17.134 (0.034) & 16.918 (0.049) & 1 \\
654.59 & 17.069 (0.029) & 18.247 (0.034) & 17.017 (0.031) & 16.861 (0.042) & 1 \\
658.55 &    \nodata     & 17.862 (0.024) & 16.745 (0.033) & 16.744 (0.029) & 2 \\
659.56 &    \nodata     & 17.831 (0.026) & 16.801 (0.025) & 16.739 (0.027) & 2 \\

661.58 &     \nodata    & 17.780 (0.028) & 16.832 (0.025) & 16.909 (0.040) & 1 \\
662.55 & 16.643 (0.026) & 17.830 (0.032) & 16.875 (0.031) & 16.791 (0.041) & 1 \\
663.54 &     \nodata    & 17.646 (0.034) & 16.800 (0.033) &     \nodata    & 1 \\
664.50 &     \nodata    & 17.649 (0.031) &     \nodata    & 16.771 (0.053) & 1 \\
666.56 & 16.449 (0.025) & 17.473 (0.025) & 16.777 (0.028) &     \nodata    & 1 \\

667.55 &     \nodata    &    \nodata     & 16.851 (0.036) & 17.116 (0.056) & 1 \\
668.50 &     \nodata    & 17.333 (0.025) & 17.078 (0.050) &     \nodata    & 1 \\
675.70 &     \nodata    & 17.967 (0.023) & 17.366 (0.031) &     \nodata    & 2 \\
676.54 &     \nodata    &    \nodata     & 17.302 (0.024) & 17.477 (0.029) & 2 \\

681.52 &     \nodata    & 18.361 (0.021) & 17.638 (0.031) &     \nodata    & 2 \\
683.51 &     \nodata    &    \nodata     & 17.723 (0.029) & 17.651 (0.037) & 2 \\
684.50 &     \nodata    & 18.577 (0.030) & 17.822 (0.036) &     \nodata    & 2 \\
\enddata
\tablenotetext{a} {Julian Date $minus$ 2,451,000.}
\tablenotetext{b} {1 = LCO 1-m.  2 = LCO 2.5-m.}
\end{deluxetable}

\begin{deluxetable}{cccccccc}
\tablecolumns{7}
\rotate
\tablewidth{0pc}
\tablecaption{Optical Photometry of SN 2000ca}
\startdata
JD\tablenotemark{a}  &  $U$  & $B$   &   $V$   &  $R$  &  $I$ &  $z$ & Telescope\tablenotemark{b} \\ \tableline \tableline
%
%
663.76 & 15.336  (0.032) &  15.870 (0.015) &  15.859 (0.012) &  15.759 (0.023) &  16.081 (0.019) &  16.022 (0.022) & 1 \\
664.77 & 15.367  (0.054) &  15.910 (0.030) &  15.850 (0.031) &  15.825 (0.019) &  16.045 (0.011) &  15.963 (0.021) & 2 \\
669.71 &      \nodata    &  15.881 (0.054) &  15.874 (0.019) &  15.779 (0.015) &  16.145 (0.028) &     \nodata     & 3 \\
672.76 & 15.882  (0.044) &  16.050 (0.011) &  15.941 (0.012) &  15.900 (0.021) &  16.392 (0.021) &     \nodata     & 2 \\
675.62 & 16.109  (0.040) &  16.218 (0.021) &  16.074 (0.012) &  16.094 (0.012) &  16.625 (0.011) &  16.396 (0.021) & 2 \\
676.69 &      \nodata    &  16.315 (0.018) &  16.098 (0.022) &  16.153 (0.012) &  16.681 (0.010) &  16.400 (0.023) & 2 \\
677.68 & 16.293  (0.028) &  16.402 (0.018) &  16.168 (0.012) &  16.236 (0.012) &  16.756 (0.011) &  16.443 (0.023) & 2 \\
681.69 &      \nodata    &  16.803 (0.024) &  16.446 (0.012) &  16.493 (0.016) &  16.930 (0.012) &  16.430 (0.026) & 2 \\
682.69 &      \nodata    &  16.979 (0.024) &  16.527 (0.015) &  16.531 (0.012) &  16.948 (0.015) &  16.446 (0.019) & 2 \\
683.70 & 17.163  (0.052) &  17.048 (0.020) &  16.549 (0.012) &  16.541 (0.012) &  16.896 (0.013) &  16.392 (0.026) & 2 \\
699.63 &      \nodata    &  18.485 (0.018) &  17.393 (0.012) &  16.947 (0.012) &  16.756 (0.010) &     \nodata     & 2 \\
705.59 &      \nodata    &  18.781 (0.023) &  17.715 (0.021) &  17.305 (0.012) &  17.113 (0.019) &     \nodata     & 2 \\
711.61 &      \nodata    &  18.913 (0.035) &  17.934 (0.032) &  17.549 (0.023) &  17.448 (0.024) &     \nodata     & 2 \\
730.53 &      \nodata    &  19.243 (0.025) &  18.455 (0.016) &  18.193 (0.022) &  18.352 (0.031) &     \nodata     & 2 \\
738.54 &      \nodata    &  19.374 (0.053) &  18.634 (0.041) &  18.441 (0.023) &  18.622 (0.036) &     \nodata     & 2 \\
745.50 &      \nodata    &  19.481 (0.039) &  18.823 (0.023) &  18.621 (0.050) &  18.832 (0.041) &     \nodata     & 2 \\
752.54 &      \nodata    &  19.536 (0.028) &  19.007 (0.020) &  18.902 (0.031) &  19.188 (0.076) &     \nodata     & 2 \\
757.51 &      \nodata    &  19.758 (0.043) &  19.137 (0.027) &  19.031 (0.036) &  19.431 (0.072) &     \nodata     & 2 \\
\enddata
\tablenotetext{a} {Julian Date $minus$ 2,451,000.}
\tablenotetext{b} {1 = ESO NTT 3.6-m.  2 = CTIO 0.9-m. 3 = LCO 1-m. }
\end{deluxetable}

\begin{deluxetable}{cccc}
\tablecolumns{4}
\tablewidth{0pc}
\tablecaption{Infrared Photometry of SN 2000ca }
\startdata
JD\tablenotemark{a}  &  $J_s$   &  $H$  & Telescope\tablenotemark{b}   \\ \tableline \tableline
%
%
663.70 &  16.392 (0.045) & 16.850 (0.048) & 1 \\
664.63 &  16.425 (0.031) & 16.711 (0.087) & 1 \\
665.73 &  16.364 (0.029) & 16.815 (0.057) & 1 \\ 
666.72 &  16.572 (0.041) & 16.841 (0.033) & 1 \\  
667.68 &  16.572 (0.037) & 16.834 (0.031) & 1 \\
668.72 &  16.588 (0.037) & 16.923 (0.040) & 1 \\
%
%
675.53 &  17.356 (0.043) & 16.964 (0.028) & 2 \\
677.74 &  17.668 (0.046) & 16.904 (0.032) & 2 \\ 
681.59 &  18.003 (0.046) & 17.002 (0.028) & 2 \\
684.54 &  18.042 (0.045) & 16.813 (0.028) & 2 \\
687.68 &  17.849 (0.046) & 16.738 (0.025) & 2 \\
688.47 &  17.774 (0.058) &    \nodata     & 2 \\
\enddata
\tablenotetext{a} {Julian Date $minus$ 2,451,000.}
\tablenotetext{b} {1 = LCO 1-m.  2 = LCO 2.5-m.}
\end{deluxetable}

\begin{deluxetable}{cccc}
\tablecolumns{4}
\tablewidth{0pc}
\tablecaption{Optical Photometry of SN 2001ba }
\startdata
JD\tablenotemark{a}  &  $B$   &   $V$   &  $I$     \\ \tableline \tableline
2030.60	& 16.551 (0.018) & 16.586 (0.027) & 16.691 (0.018) \\
2031.56	& 16.465 (0.020) & 16.542 (0.029) & 16.709 (0.017) \\
2032.59	& 16.461 (0.015) & 16.498 (0.023) & 16.679 (0.019) \\
2033.58	& 16.451 (0.019) & 16.471 (0.027) & 16.686 (0.021) \\
2034.59	& 16.445 (0.016) & 16.465 (0.031) & 16.746 (0.024) \\

2037.58	& 16.558 (0.022) & 16.524 (0.028) & 16.872 (0.033) \\
2038.56	& 16.679 (0.069) & 16.509 (0.061) & 16.796 (0.129) \\
2040.61	& 16.701 (0.033) & 16.610 (0.033) & 17.041 (0.040) \\
2042.59	& 16.777 (0.034) & 16.674 (0.034) & 17.119 (0.046) \\
2045.71	& 17.112 (0.051) & 16.734 (0.059) & 17.304 (0.122) \\

2046.57	& 17.171 (0.021) & 16.847 (0.032) & 17.508 (0.037) \\
2047.57	& 17.233 (0.088) &    \nodata     & 17.455 (0.136) \\
2048.66	& 17.431 (0.030) & 16.951 (0.037) & 17.566 (0.066) \\
2049.54	& 17.462 (0.023) & 17.069 (0.026) & 17.567 (0.037) \\
2051.53	& 17.745 (0.031) & 17.138 (0.029) &      \nodata   \\

2053.50	& 17.934 (0.023) & 17.260 (0.028) & 17.494 (0.035) \\
2054.54	& 18.083 (0.026) & 17.339 (0.032) &      \nodata    \\
2055.55	& 18.208 (0.036) & 17.373 (0.033) & 17.467 (0.040) \\
2057.50	& 18.389 (0.027) & 17.486 (0.033) & 17.405 (0.031) \\
2059.52	& 18.482 (0.047) & 17.629 (0.049) & 17.377 (0.049) \\

2060.52	&     \nodata    & 17.640 (0.075) & 17.213 (0.074) \\
2061.61	& 18.884 (0.071) & 17.648 (0.044) & 17.351 (0.038) \\
2063.53	& 18.908 (0.044) & 17.986 (0.039) &     \nodata    \\
2064.50	& 19.005 (0.083) & 17.872 (0.062) & 17.528 (0.071) \\
2066.50	&     \nodata    & 17.974 (0.073) & 17.406 (0.064) \\

2067.51	& 19.240 (0.055) & 18.101 (0.045) & 17.527 (0.068) \\
2068.50	& 19.292 (0.054) & 18.093 (0.044) & 17.540 (0.056) \\
2069.50	& 19.349 (0.061) & 18.275 (0.046) & 17.627 (0.060) \\
2070.49	& 19.290 (0.071) & 18.363 (0.057) & 17.721 (0.063) \\
2072.48	&     \nodata    & 18.406 (0.079) &     \nodata     \\

2075.49	& 19.706 (0.128) &     \nodata    &     \nodata     \\
2077.48	&     \nodata    &     \nodata    & 18.042 (0.085) \\
2079.54	&     \nodata    & 18.632 (0.073) & 18.133 (0.105) \\
2080.51	&     \nodata    &     \nodata    & 18.261 (0.075) \\
2082.49	& 19.626 (0.078) & 18.846 (0.065) & 18.339 (0.086) \\

2083.51	&     \nodata    & 18.810 (0.114) &     \nodata     \\
2086.48	&     \nodata    & 18.698 (0.068) &     \nodata     \\
2087.49	&     \nodata    &     \nodata    & 18.317 (0.077) \\
2088.48	&     \nodata    &     \nodata    & 18.534 (0.109) \\
2090.46	& 20.140 (0.222) &     \nodata    &     \nodata     \\

2091.48	&     \nodata    & 18.948 (0.098) & 18.534 (0.086) \\
\enddata
\tablenotetext{a} {Julian Date $minus$ 2,450,000.}
\end{deluxetable}

\begin{deluxetable}{ccccc}
\tablecolumns{5}
\tablewidth{0pc}
\tablecaption{Infrared Photometry of SN 2001ba }
\startdata
JD\tablenotemark{a}  &  $J_s$   &  $H$  & $K_s$  & Telescope\tablenotemark{b}   \\ \tableline \tableline
2028.58 &   17.027 (0.033) &      \nodata      &     \nodata     & 1 \\          
2029.56 &   16.998 (0.023) &   17.240 (0.040)  &  17.187 (0.062) & 1 \\
2030.53 &      \nodata     &   17.251 (0.038)  &     \nodata     & 1 \\
2031.54 &   16.992 (0.018) &   17.332 (0.045)  &  17.161 (0.081) & 1 \\
2032.58 &   16.994 (0.029) &   17.310 (0.057)  &  17.061 (0.064) & 1 \\
2044.57 &   18.362 (0.053) &   17.920 (0.056)  &     \nodata     & 1 \\
2045.52 &   18.671 (0.045) &   17.777 (0.077)  &  17.441 (0.029) & 1 \\
2047.53 &   18.977 (0.082) &   17.941 (0.075)  &  17.525 (0.087) & 1 \\
2048.48 &   19.115 (0.136) &      \nodata      &     \nodata     & 1 \\
2049.54 &   18.742 (0.095) &   17.855 (0.108)  &  17.422 (0.093) & 1 \\
2052.55 &   18.872 (0.076) &   17.771 (0.048)  &  17.554 (0.116) & 1 \\
2055.54 &   18.525 (0.061) &   17.837 (0.107)  &  17.246 (0.073) & 1 \\
2057.54 &   18.514 (0.052) &   17.437 (0.045)  &  17.389 (0.079) & 1 \\
2061.56 &   18.314 (0.031) &   17.448 (0.056)  &     \nodata     & 2 \\
2068.46 &   18.006 (0.029) &   17.485 (0.029)  &     \nodata     & 2 \\
2069.46 &   18.021 (0.035) &   17.516 (0.027)  &     \nodata     & 2 \\
\enddata
\tablenotetext{a} {Julian Date $minus$ 2,450,000.}
\tablenotetext{b} {1 = LCO 1-m.  2 = LCO 2.5-m.}
\end{deluxetable}

\clearpage

\pagestyle{empty}
\begin{deluxetable}{ccccccccccccccccc}
\tablecolumns{17}
\tablewidth{0pc}
\tabletypesize{\scriptsize}
\rotate
\tablecaption{K-corrections for Type Ia Supernovae in LCO Near Infrared Bands\tablenotemark{a}}
\startdata
$t$\tablenotemark{b} $\; \backslash\; z$ & 0.005 & 0.010 & 0.015 & 0.020 & 0.025 & 0.030 & 0.035 & 0.040 & 0.045 & 0.050 & 
0.060 & 0.070 & 0.080 & 0.090 & 0.100 & 0.150 \\ \tableline \tableline
        &       &        &        &        &        &        &        &  $J_s$:  &        &        &        &        &        &        &        &        \\
$-$8.57 & $-$0.012 & $-$0.023 & $-$0.034 & $-$0.044 & $-$0.055 & $-$0.066 & $-$0.078 & $-$0.090 & $-$0.103 & $-$0.116 & $-$0.141 & $-$0.164 & $-$0.185 & $-$0.204 & $-$0.221 & $-$0.321 \\
\phs1.40   & $-$0.007 & $-$0.015 & $-$0.021 & $-$0.027 & $-$0.034 & $-$0.041 & $-$0.047 & $-$0.053 & $-$0.058 & $-$0.062 & $-$0.066 & $-$0.066 & $-$0.064 & $-$0.061 & $-$0.061 & $-$0.119 \\
\phs4.51   & $-$0.003 & $-$0.007 & $-$0.010 & $-$0.012 & $-$0.014 & $-$0.018 & $-$0.023 & $-$0.026 & $-$0.028 & $-$0.030 & $-$0.031 & $-$0.027 & $-$0.022 & $-$0.017 & $-$0.017 & $-$0.090 \\
\phs8.38   & $-$0.008 & $-$0.016 & $-$0.023 & $-$0.030 & $-$0.037 & $-$0.042 & $-$0.048 & $-$0.052 & $-$0.055 & $-$0.058 & $-$0.058 & $-$0.056 & $-$0.057 & $-$0.062 & $-$0.077 & $-$0.241 \\
15.39   & $-$0.013 & $-$0.030 & $-$0.045 & $-$0.062 & $-$0.082 & $-$0.101 & $-$0.122 & $-$0.146 & $-$0.171 & $-$0.194 & $-$0.239 & $-$0.286 & $-$0.339 & $-$0.407 & $-$0.485 & $-$0.878 \\
19.42   & $-$0.019 & $-$0.039 & $-$0.061 & $-$0.084 & $-$0.109 & $-$0.133 & $-$0.158 & $-$0.184 & $-$0.213 & $-$0.241 & $-$0.300 & $-$0.366 & $-$0.442 & $-$0.533 & $-$0.625 & $-$1.058 \\
22.40   & $-$0.023 & $-$0.047 & $-$0.072 & $-$0.099 & $-$0.127 & $-$0.155 & $-$0.184 & $-$0.216 & $-$0.249 & $-$0.283 & $-$0.354 & $-$0.434 & $-$0.523 & $-$0.623 & $-$0.723 & $-$1.164 \\
27.46   & $-$0.034 & $-$0.069 & $-$0.104 & $-$0.137 & $-$0.173 & $-$0.206 & $-$0.242 & $-$0.280 & $-$0.321 & $-$0.361 & $-$0.448 & $-$0.540 & $-$0.642 & $-$0.752 & $-$0.862 & $-$1.319 \\
31.43   & $-$0.028 & $-$0.058 & $-$0.089 & $-$0.121 & $-$0.153 & $-$0.187 & $-$0.222 & $-$0.260 & $-$0.300 & $-$0.343 & $-$0.436 & $-$0.540 & $-$0.648 & $-$0.757 & $-$0.871 & $-$1.328 \\
41.42   & $-$0.028 & $-$0.064 & $-$0.105 & $-$0.149 & $-$0.196 & $-$0.244 & $-$0.294 & $-$0.347 & $-$0.403 & $-$0.459 & $-$0.561 & $-$0.656 & $-$0.756 & $-$0.861 & $-$0.973 & \nodata  \\
\tableline
     &       &        &        &        &        &        &        &  $H$:   &        &        &        &        &        &        &        &        \\
$-$8.57 & $-$0.008 & $-$0.016 & $-$0.025 & $-$0.033 & $-$0.042 & $-$0.051 & $-$0.060 & $-$0.070 & $-$0.079 & $-$0.088 & $-$0.106 & $-$0.125 & $-$0.144 & $-$0.166 & $-$0.191 & $-$0.373 \\
\phs0.39   & $-$0.004 & $-$0.007 & $-$0.010 &  \phs0.007 & \phs0.007 & \phs0.016 & \phs0.004 & $-$0.001 & $-$0.006 & $-$0.013 & $-$0.030 & $-$0.049 & $-$0.077 & $-$0.108 & $-$0.146 & $-$0.397 \\
\phs1.40   & \phs0.002 & \phs0.005 & \phs0.008 & \phs0.011 & \phs0.013 & \phs0.013 & \phs0.011 & \phs0.007 & \phs0.002 & $-$0.005 & $-$0.021 & $-$0.041 & $-$0.069 & $-$0.102 & $-$0.141 & $-$0.400 \\
\phs4.51   & \phs0.008 & \phs0.017 & \phs0.028 & \phs0.038 & \phs0.048 & \phs0.054 & \phs0.057 & \phs0.061 & \phs0.063 & \phs0.065 & \phs0.061 & \phs0.049 & \phs0.024 & $-$0.007 & $-$0.045 & $-$0.290 \\
\phs8.38   & \phs0.004 & \phs0.011 & \phs0.021 & \phs0.034 & \phs0.047 & \phs0.060 & \phs0.073 & \phs0.087 & \phs0.102 & \phs0.115 & \phs0.133 & \phs0.136 & \phs0.122 & \phs0.098 & \phs0.072 & $-$0.071 \\
15.39   & \phs0.015 & \phs0.034 & \phs0.056 & \phs0.082 & \phs0.109 & \phs0.139 & \phs0.170 & \phs0.203 & \phs0.234 & \phs0.263 & \phs0.315 & \phs0.354 & \phs0.384 & \phs0.404 & \phs0.420 & \phs0.549 \\
19.42   & \phs0.015 & \phs0.035 & \phs0.062 & \phs0.093 & \phs0.126 & \phs0.158 & \phs0.189 & \phs0.218 & \phs0.244 & \phs0.267 & \phs0.306 & \phs0.338 & \phs0.367 & \phs0.393 & \phs0.420 & \phs0.622 \\
22.40   & \phs0.017 & \phs0.039 & \phs0.065 & \phs0.096 & \phs0.129 & \phs0.162 & \phs0.194 & \phs0.222 & \phs0.249 & \phs0.273 & \phs0.312 & \phs0.346 & \phs0.376 & \phs0.406 & \phs0.436 & \phs0.652 \\
27.46   & \phs0.011 & \phs0.026 & \phs0.045 & \phs0.069 & \phs0.098 & \phs0.122 & \phs0.144 & \phs0.163 & \phs0.180 & \phs0.196 & \phs0.225 & \phs0.257 & \phs0.281 & \phs0.304 & \phs0.338 & \phs0.607 \\
31.43   & $-$0.002 & \phs0.002 & \phs0.012 & \phs0.024 & \phs0.038 &  \phs0.051 & \phs0.061 & \phs0.068 & \phs0.075 & \phs0.080 & \phs0.089 & \phs0.097 & \phs0.104 & \phs0.117 & \phs0.145 & \phs0.344 \\
41.42   & $-$0.014 & $-$0.023 & $-$0.026 & $-$0.025 & $-$0.022 & $-$0.018 & $-$0.015 & $-$0.012 & $-$0.010 & $-$0.006 & \phs0.001 & \phs0.010 & \phs0.023 & \phs0.046 & \phs0.082 & \phs0.338 \\
\tableline
        &       &        &        &        &        &        &        &  $K_s$:   &        &        &        &        &        &        &        &        \\
$-$8.57 & $-$0.012 & $-$0.024 & $-$0.036 & $-$0.048 & $-$0.058 & $-$0.070 & $-$0.082 & $-$0.095 & $-$0.106 & $-$0.117 & $-$0.143 & $-$0.170 & $-$0.195 & $-$0.221 & $-$0.246 & $-$0.367 \\
\phs0.39   & $-$0.021 & $-$0.047 & $-$0.074 & $-$0.102 & $-$0.130 & $-$0.158 & $-$0.189 & $-$0.219 & $-$0.248 & $-$0.275 & $-$0.330 & $-$0.382 & $-$0.433 & $-$0.475 & $-$0.517 & $-$0.706 \\
\phs1.40   & $-$0.025 & $-$0.053 & $-$0.082 & $-$0.112 & $-$0.143 & $-$0.175 & $-$0.208 & $-$0.239 & $-$0.269 & $-$0.300 & $-$0.362 & $-$0.421 & $-$0.477 & $-$0.530 & $-$0.580 & $-$0.813 \\
\phs4.51   & $-$0.033 & $-$0.072 & $-$0.113 & $-$0.155 & $-$0.194 & $-$0.231 & $-$0.269 & $-$0.308 & $-$0.344 & $-$0.374 & $-$0.435 & $-$0.492 & $-$0.546 & $-$0.601 & $-$0.649 & $-$0.851 \\
\phs8.38   & $-$0.036 & $-$0.079 & $-$0.122 & $-$0.162 & $-$0.202 & $-$0.240 & $-$0.279 & $-$0.315 & $-$0.350 & $-$0.387 & $-$0.450 & $-$0.504 & $-$0.553 & $-$0.593 & $-$0.629 & $-$0.797 \\
15.39   & $-$0.033 & $-$0.066 & $-$0.097 & $-$0.127 & $-$0.157 & $-$0.187 & $-$0.216 & $-$0.241 & $-$0.265 & $-$0.287 & $-$0.329 & $-$0.365 & $-$0.398 & $-$0.433 & $-$0.474 & $-$0.696 \\
22.40   & $-$0.018 & $-$0.035 & $-$0.051 & $-$0.066 & $-$0.080 & $-$0.092 & $-$0.099 & $-$0.105 & $-$0.106 & $-$0.108 & $-$0.117 & $-$0.130 & $-$0.148 & $-$0.179 & $-$0.216 & $-$0.534 \\
27.46   & $-$0.002 & $-$0.006 & $-$0.015 & $-$0.021 & $-$0.027 & $-$0.024 & $-$0.011 & $-$0.003 & \phs0.004 & \phs0.008 & \phs0.026 & \phs0.016 & $-$0.015 & $-$0.058 & $-$0.099 & $-$0.460 \\
31.43   & $-$0.004 & $-$0.012 & $-$0.024 & $-$0.036 & $-$0.043 & $-$0.043 & $-$0.036 & $-$0.025 & $-$0.015 & $-$0.006 & \phs0.005 & $-$0.013 & $-$0.043 & $-$0.071 & $-$0.100 &$-$0.581 \\
41.42   & $-$0.004 & $-$0.013 & $-$0.024 & $-$0.036 & $-$0.048 & $-$0.047 & $-$0.036 & $-$0.023 & $-$0.012 & $-$0.006 & \phs0.007 & \phs0.000 & $-$0.020 & $-$0.028 & $-$0.036 & $-$0.373 \\
 
\enddata
\tablenotetext{a} {The following corrections are {\em subtracted} from the photometric data to correct them
to the observer's frame (redshift $z$ = 0.00).}
\tablenotetext{b} {Time $t$ is the number of days since $B$-band maximum.}
\end{deluxetable}

\clearpage

\pagestyle{plain}

\begin{deluxetable}{cccc}
\tablecolumns{4} 
\tablewidth{0pc}
\tablecaption{Fitting Supernova Light Curves at Maximum\tablenotemark{a}}
\startdata
Coeff &     $\Delta J$    & $\Delta H$  &  $\Delta K$  \\ \tableline \tableline
$a_0$ &  0.080          &  0.050         &  0.042         \\
$a_1$ &  \phs 0.5104699E$-$01 &  \phs 0.2509234E$-$01  &   \phs 0.2728437E$-$01 \\
$a_2$ &  \phs 0.7064257E$-$02 &  \phs 0.1852107E$-$02 &  \phs 0.3194500E$-$02  \\
$a_3$ & $-$0.2579062E$-$03 & $-$0.3557824E$-$03 & $-$0.4139377E$-$03 \\
\enddata
\tablenotetext{a} {The differential magnitudes with respect
to maximum are of the form: $\Delta$ mag = $a_0$ + $\Sigma$($a_i \times t^i$),
where $t$ is the number of days since $B$-band maximum, scaled
according to the mean of the $B$- and $V$-band inverse stretch factors
 $s^{-1}$ from Jha (2000, Fig. 3.8) and divided by (1 + $z$). The fits 
are valid from $-12$ to +10 d of ``stretched time''.}
\end{deluxetable}

\begin{deluxetable}{lccc}
\tablecolumns{4}
\tablewidth{0pc}
\tablecaption{$V$ minus Infrared Loci for Slowly Declining Type Ia Supernovae\tablenotemark{a}}
\startdata
Parameter $\backslash$ color index  &  $V-J$               & $V-H$                &  $V-K$ \\ \tableline \tableline
$a_0$              &  $-$0.902(0.005)     &  $-$1.194            &  $-$0.971           \\
$a_1$              &  $-$0.07136(0.00104) &  $-$0.4793682E$-$01  &  $-$0.4089032E$-$01 \\
$a_2$              &                      &  \phs 0.4647584E$-$02     &  \phs 0.3613656E$-$02  \\
$a_3$              &                      &  \phs 0.8813858E$-$04     &  \phs 0.6178713E$-$04    \\
$a_4$              &                      & $-$0.5205402E$-$05   &  $-$0.3648954E$-$05 \\
Range\tablenotemark{b}   & [$-$8,+9.5]          & [$-$8,+8.5]          & [$-8$,+27]          \\
{\sc rms} residual & $\pm$ 0.067          &  $\pm$ 0.062         & $\pm$ 0.138         \\
$\chi^2_{\nu}$     & 3.5                  &  1.9                 & 4.9                 \\
\enddata
\tablenotetext{a} {The range of $\Delta$m$_{15}$($B$) is 0.81 to 1.00.
The fits are of the form: $V \; - \; $ [$J,H,K$] = $a_0$ + $\Sigma$($a_i \times t^i$),
where $t$ is the number of days since $B$-band maximum, corrected for time dilation
but not stretch.}
\tablenotetext{b} {The range in days with respect to $B$-band maximum for which the fit is
most valid.}
\end{deluxetable}

\begin{deluxetable}{lccc}
\tablecolumns{4}
\tablewidth{0pc}
\tablecaption{$V$ minus Infrared Loci for Type Ia Supernovae of Mid-Range Decline Rates\tablenotemark{a}}
\startdata
Parameter $\backslash$ color index  & $V-H$                &  $V-K$ \\ \tableline \tableline
$a_0$              &  $-$0.951            &  $-$0.741           \\
$a_1$              &  $-$0.5383673E$-$01  &  $-$0.5071254E$-$01 \\
$a_2$              &  \phs0.5913753E$-$02     &  \phs 0.3281215E$-$02  \\
$a_3$              &  \phs0.1155147E$-$03     &  \phs 0.2158419E$-$03    \\
$a_4$              & $-$0.7144163E$-$05   &  $-$0.8056540E$-$05 \\
Range\tablenotemark{b}   & [$-$9,+27]          & [$-9$,+27]          \\
{\sc rms} residual &  $\pm$ 0.093         & $\pm$ 0.105         \\
$\chi^2_{\nu}$     &  4.2                 & 3.9                 \\
\enddata
\tablenotetext{a} {Krisciunas et al. (2000, Table 9) gave hinged two-line fits to
the color curves.  Actual data are redder at $t$ = +6 d than the two-line fits.
Here we give fourth order polynomial fits to the same extinction-corrected data.
The fits are of the form: $V \; - \; $ [$J,H,K$] = $a_0$ + $\Sigma$($a_i \times t^i$),
where $t$ is the number of days since $B$-band maximum.}
\tablenotetext{b} {The range in days with respect to $B$-band maximum for which the fit is
most valid.}
\end{deluxetable}

\clearpage


\figcaption[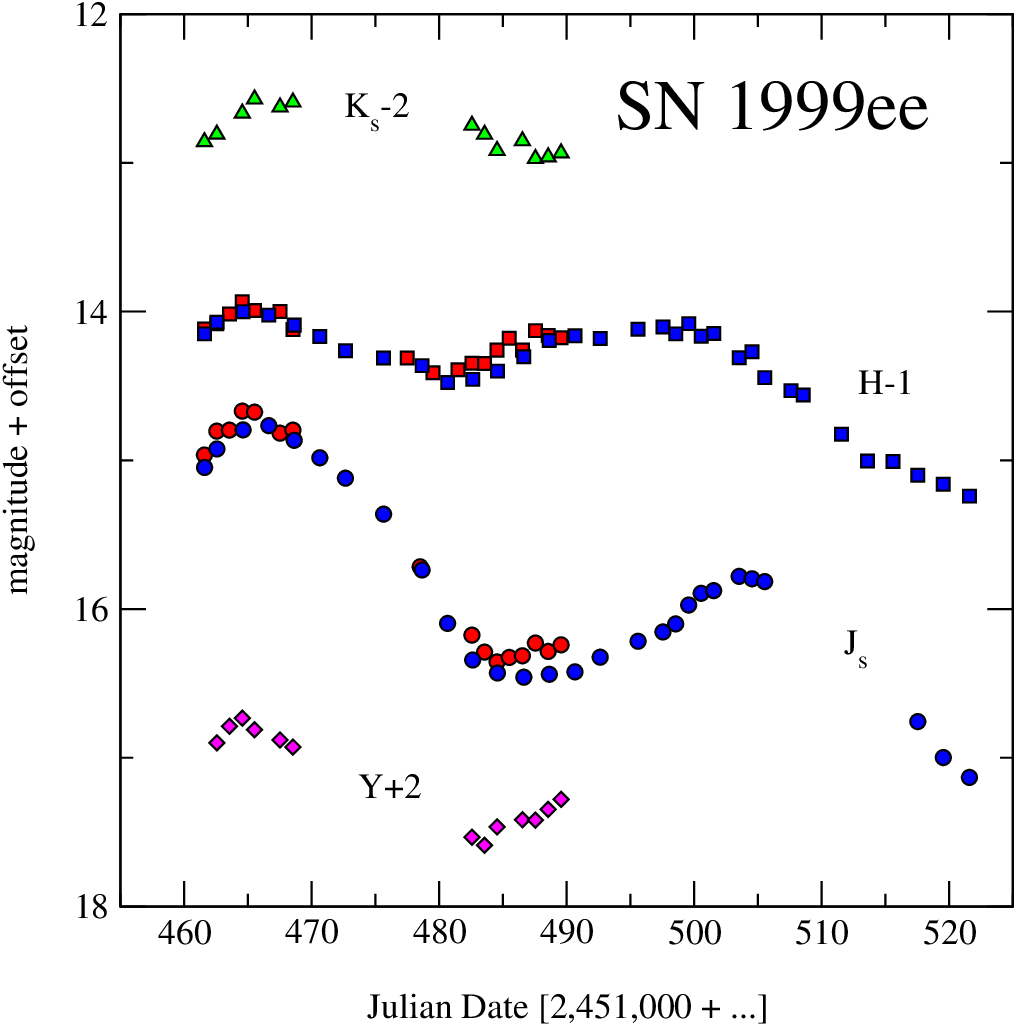] {Near infrared photometry of SN 1999ee.
For $J_s$ and $H$ the Las Campanas data are red filled in symbols,
while the YALO data are filled in with blue.  The YALO data include 
filter-based corrections to place them on the system of Persson
et al. (1998) used at Las Campanas Observatory.
\label{99ee_yjhk}
}

\figcaption[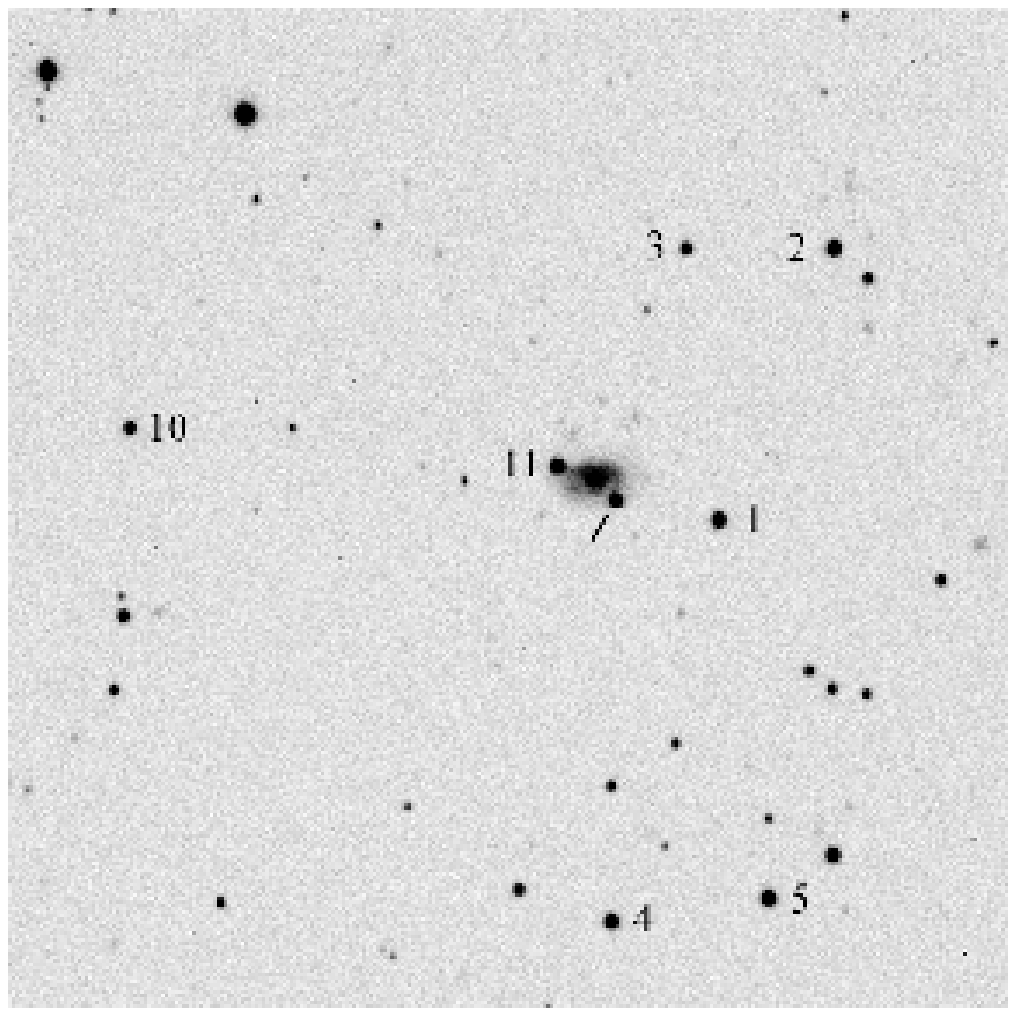] {Finding chart for SN 2000bh. North is up and east to the
left.  The image is 6.5 arcmin on a side.  It is a V-band image obtained on the
CTIO 0.9-m telescope.  The SN is indicated by the short line.
\label{sn2000bh}
}

\figcaption[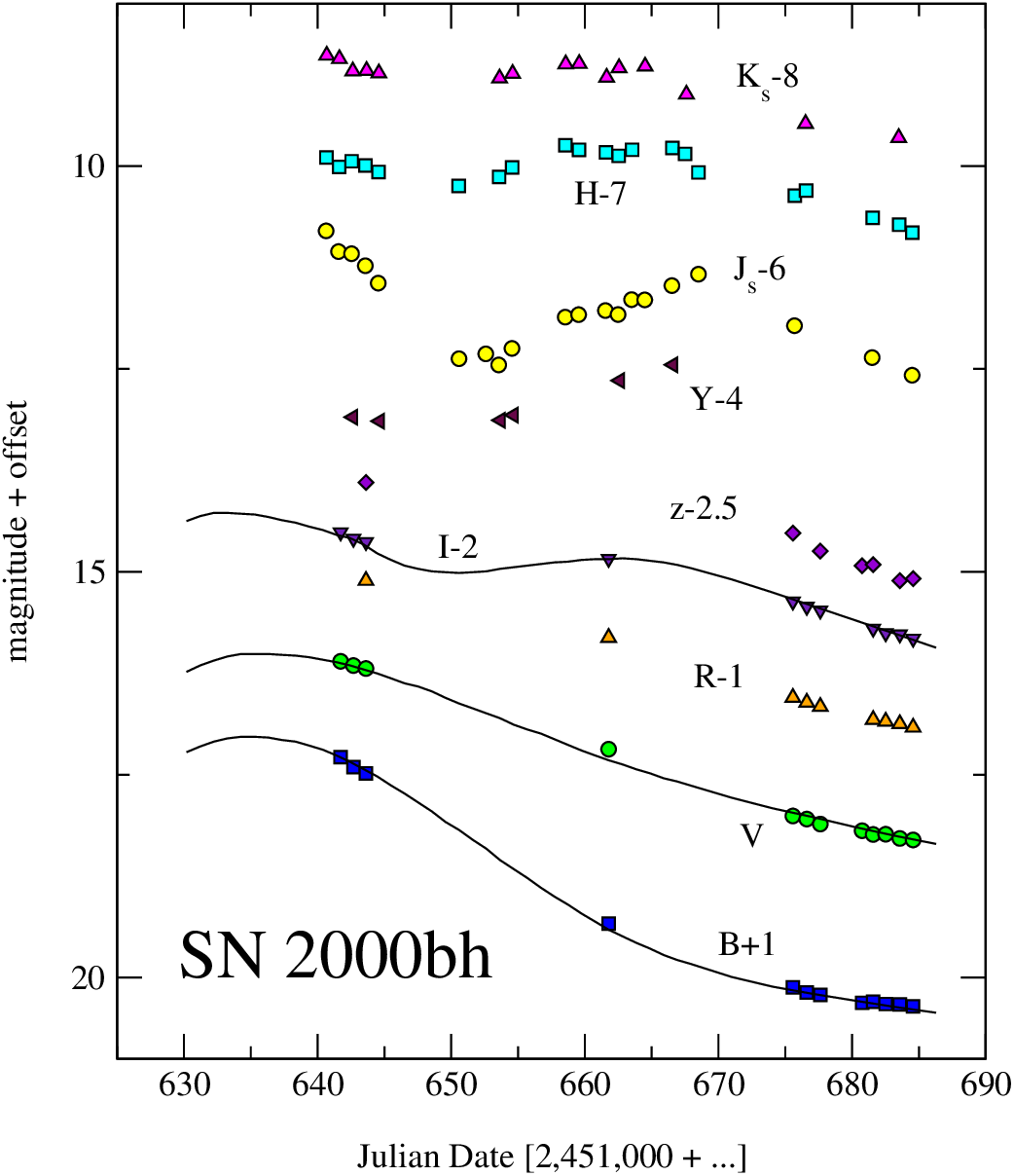] {Optical and infrared photometry of SN 2000bh.
We have added $BVI$ fits derived using the $\Delta$m$_{15}$ method of Phillips
et al. (1999). \label{00bh_opt_ir}
}

\figcaption[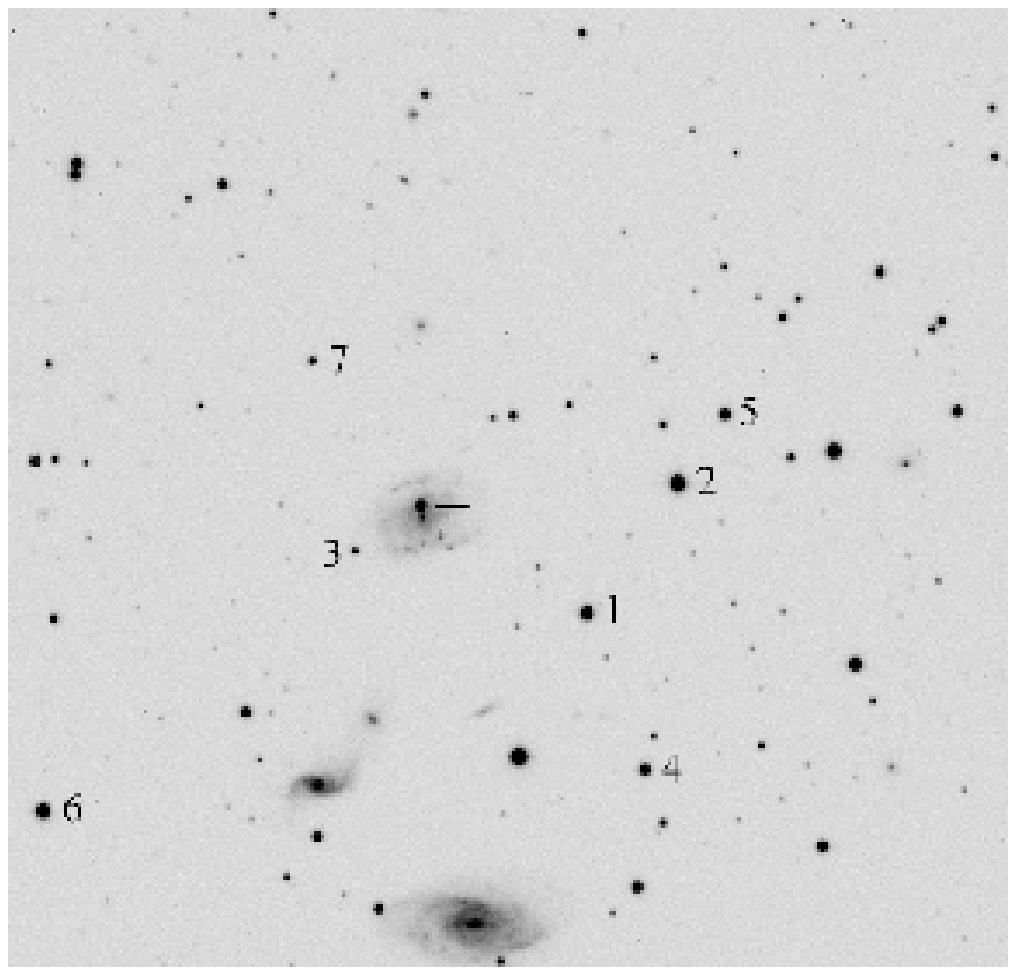] {Similar to Fig. \ref{sn2000bh}, but for SN 2000ca.
\label{sn2000ca}
}

\figcaption[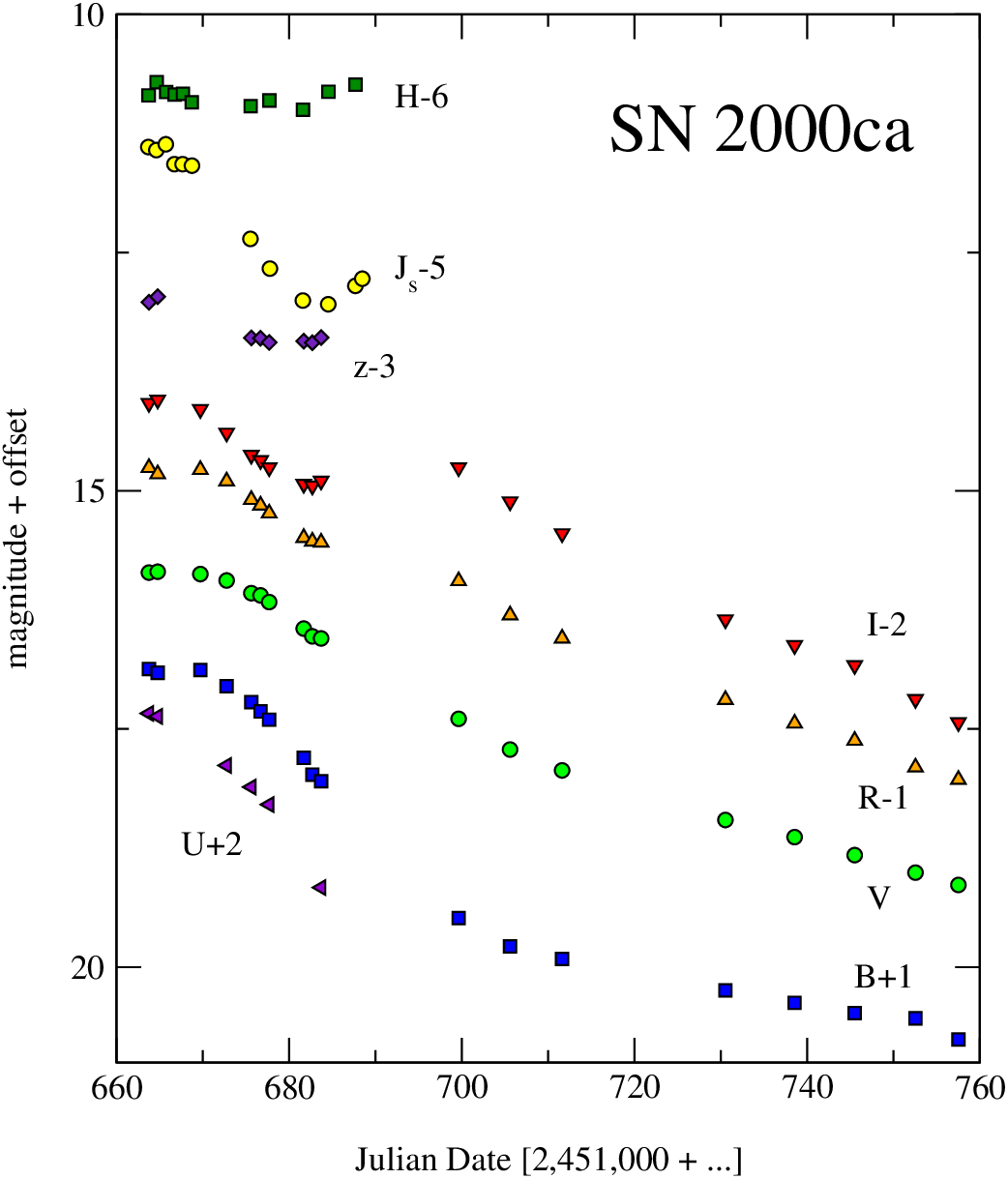] {Optical and infrared photometry of SN 2000ca.
\label{00ca_opt_ir}
}

\figcaption[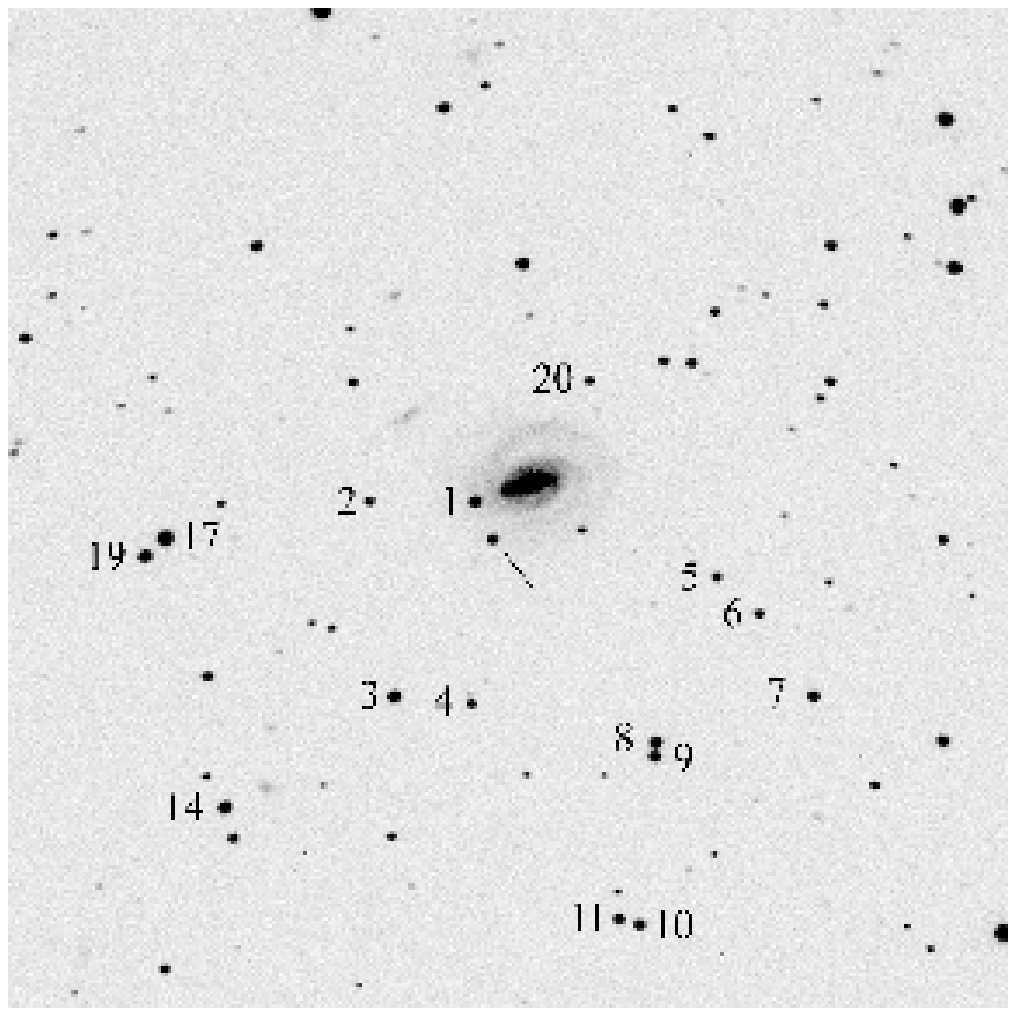] {Similar to Fig. \ref{sn2000bh}, 
but for SN 2001ba. This V-band image was obtained with the Las Campanas 2.5-m
telescope. 
\label{sn2001ba}
}

\figcaption[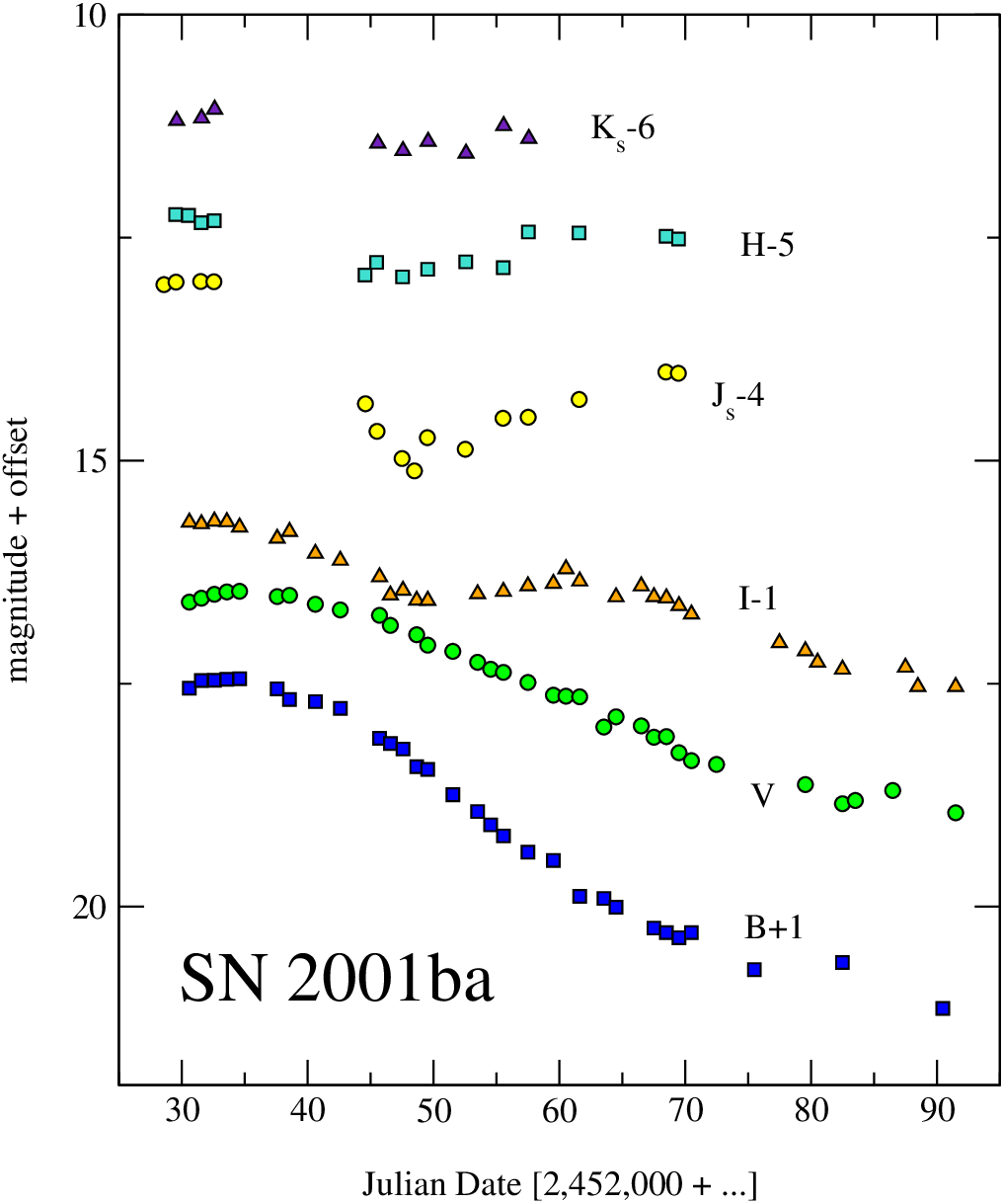] {Optical and infrared photometry of SN 2001ba.
\label{01ba_opt_ir}
}

\figcaption[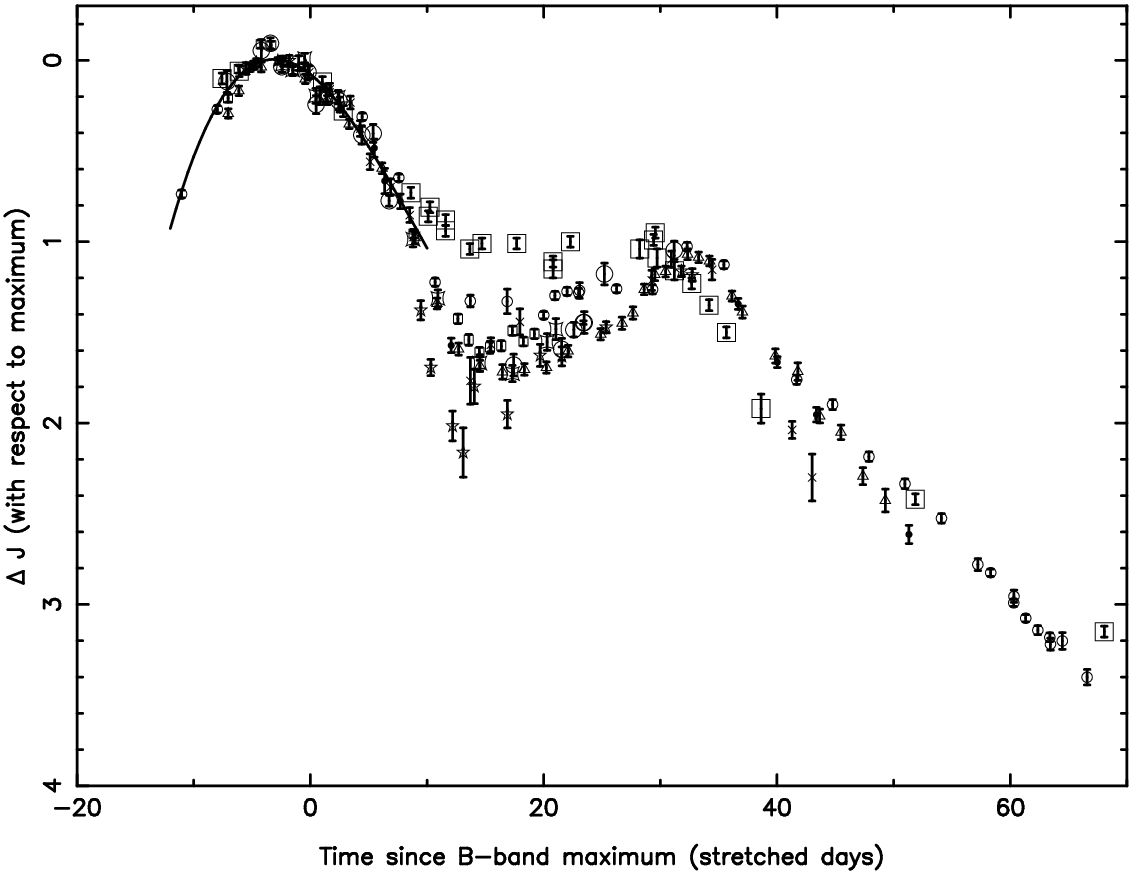] {$J$-band light curves of 8 Type Ia SNe,
normalized to maximum brightness and stretched in the time axis according
to inverse stretch factors derived from Fig. 3.8 of Jha (2002).  
We have also take out the time dilation.  The
key to the symbols is: 1980N (solid dots), 
1986G (large open squares), 1998bu (larger open circles), 1999aw (X's),
1999ee (LCO data = small open squares, corrected YALO data = triangles),
2000ca (pinched in squares), 2001ba (stars),
2001el (smaller open circles). \label{j_light_curves}
}

\figcaption[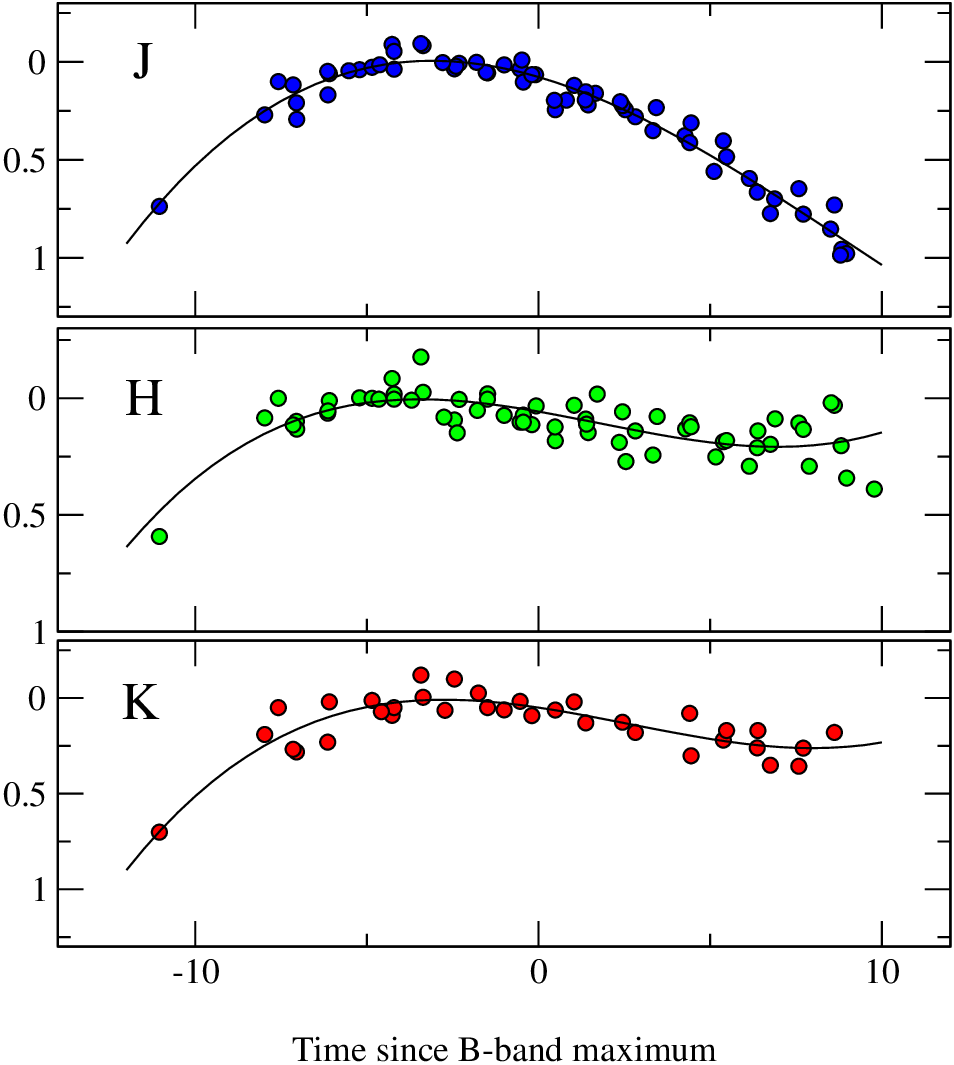] {$JH$ maxima of 8 Type Ia SNe
and $K$-band maxima of 6 Type Ia SNe.  The time axis is ``stretched
days'' (see Fig. \ref{j_light_curves} and the text).  We also give third
order polynomial fits.  \label{jhk_maxima}
}

\figcaption[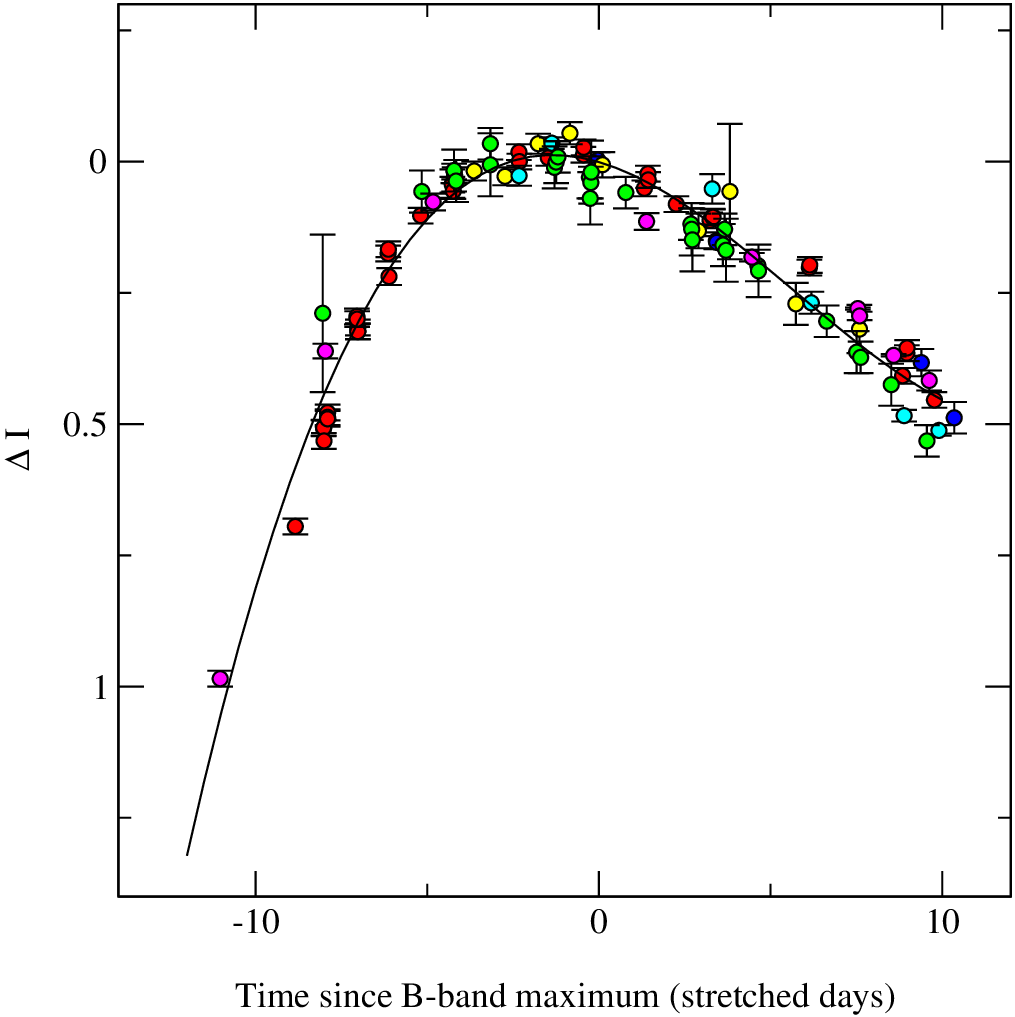] {$I$-band data of SNe 1998bu, 1999aw, 1999ee,
2000ca, 2001ba, and 2001el.  The data are stretched in the time axis
in a manner similar to Figs. \ref{j_light_curves} and \ref{jhk_maxima}.
\label{i_stretch}
}

\figcaption[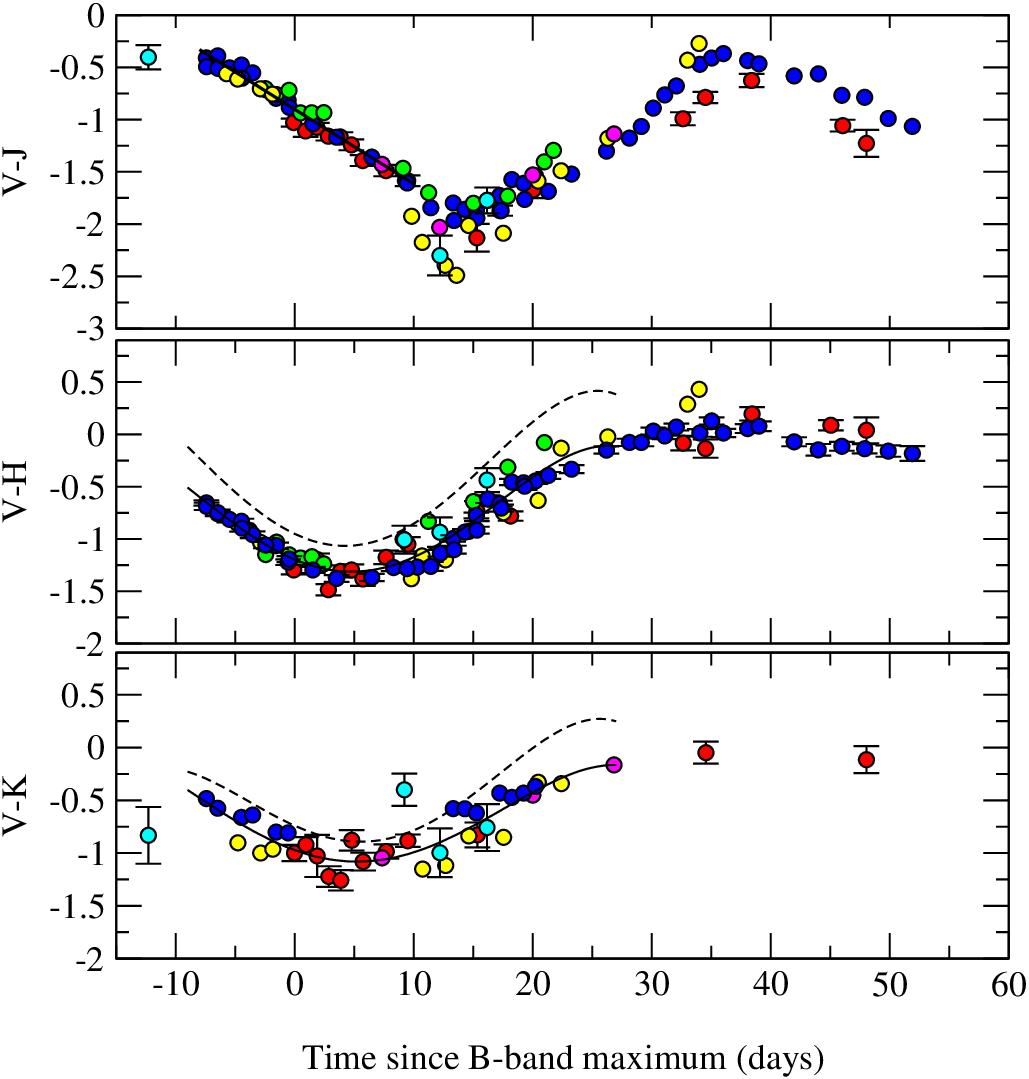] {$V \; minus$ IR colors of Type Ia SNe with decline
rates 0.81 $\leq$ $\Delta$m$_{15}$($B$) $\leq$ 1.00. The data are
K-corrected, and have been corrected for extinction due to dust and time
dilation, but not stretch.  The solid line in the top plot shows the
range of uniformity of $V-J$.  The dashed lines in the bottom two plots
are the loci for mid-range decliners (Krisciunas et al. 2000, 2003). The
solid lines in the bottom two plots are fourth order polynomial fits to
the data from $-$8 to +27 days with respect to $B$-band maximum.  On the
whole the random errors of the photometry are on the order of the size of
the points, or smaller.  In the cases of SNe~1999gp and 1999aw we show
the considerably larger error bars.  The points are color coded as
follows: SN~1999aa (magenta); 1999aw (red), 1999ee (blue), 1999gp (cyan),
2000ca (green), and 2001ba (yellow). \label{vjhk}
}

\figcaption[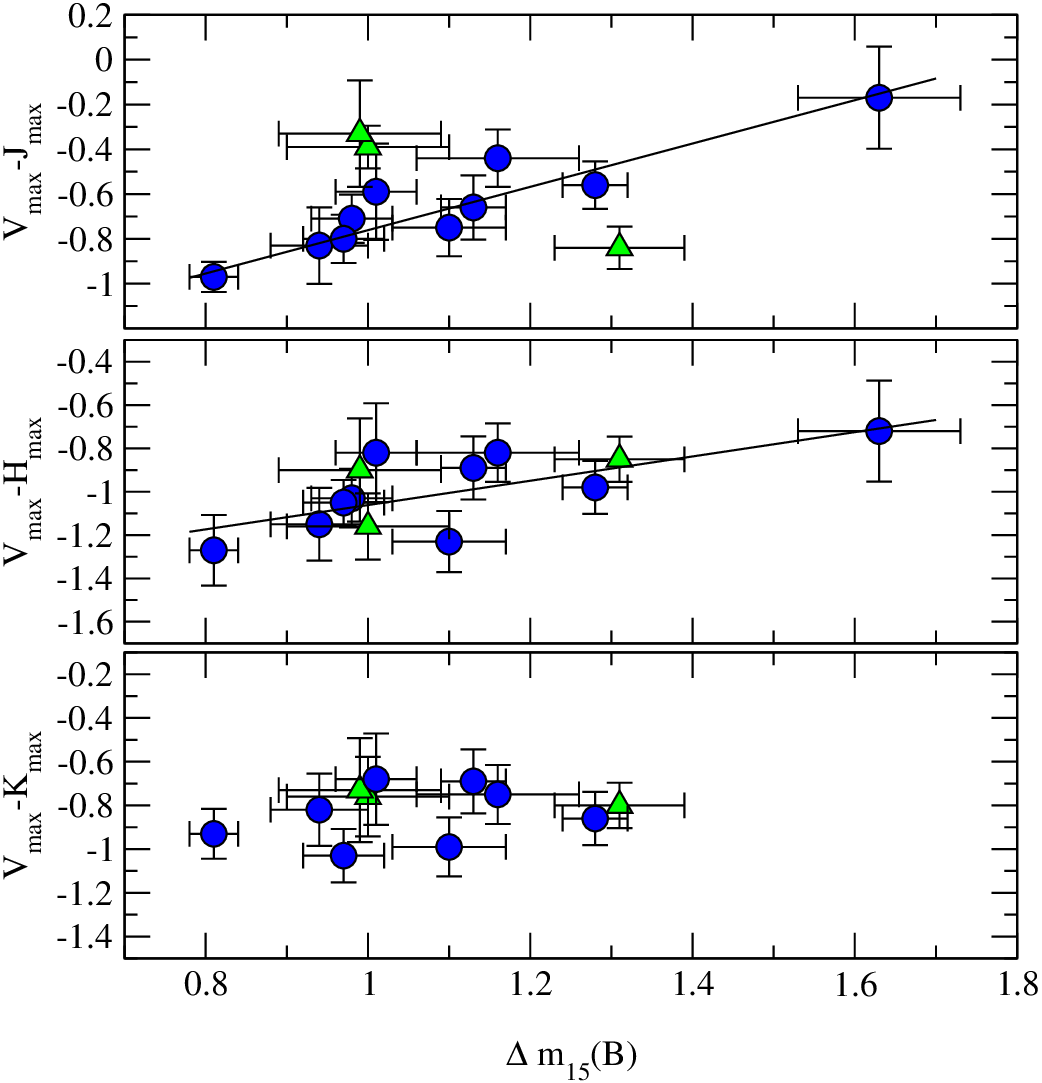] {Dereddened colors of Type Ia SNe
vs. the decline rate parameter $\Delta$m$_{15}$($B$).  The (green)
triangles correspond to objects which have few infrared observations
in the [$-$12,+10] day window, so their IR maxima are not as well 
determined; these objects were not used for the regression lines
shown. \label{vjhkmax}
}

\figcaption[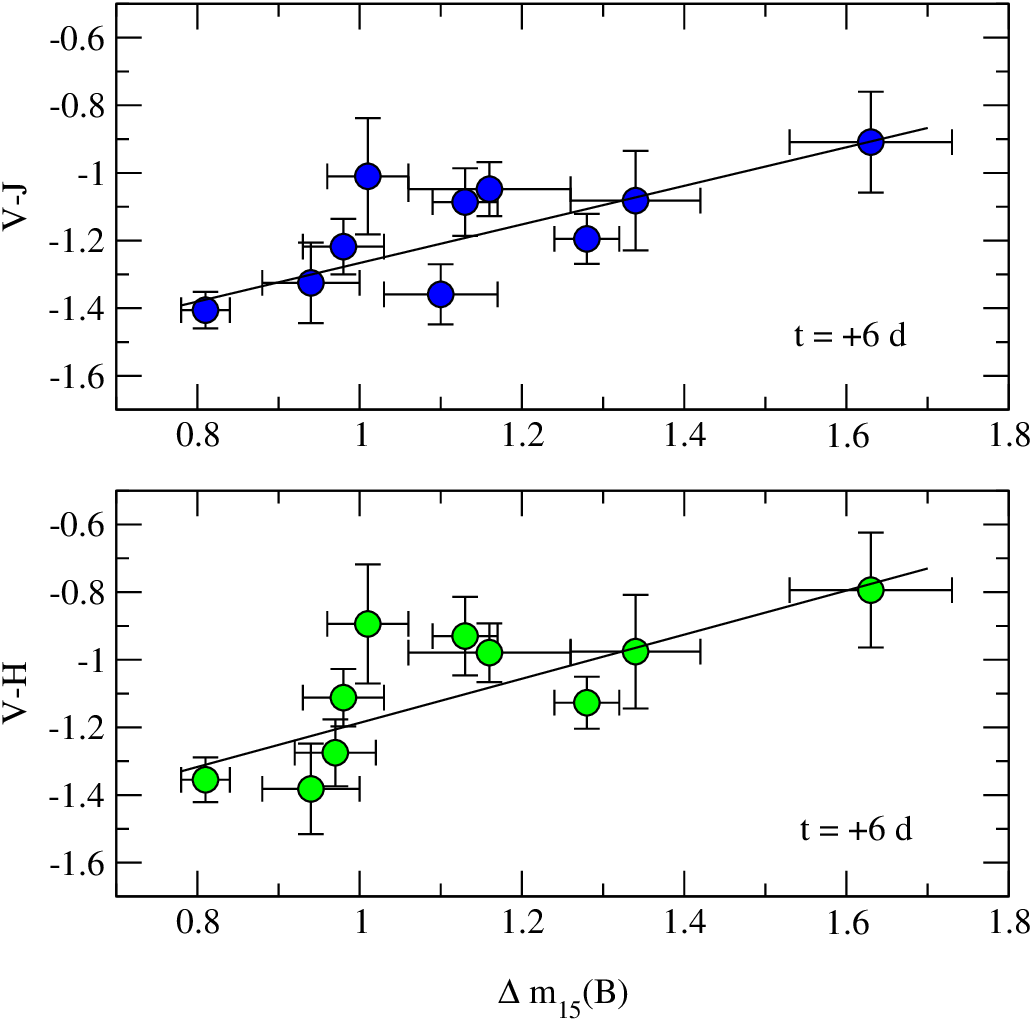] {Dereddened $V-J$ and $V-H$ colors of Type Ia SNe
at 6 days after the time of the $B$-band maxima, vs. the decline
rate parameter $\Delta$m$_{15}$($B$).  \label{vjh6}
}

\figcaption[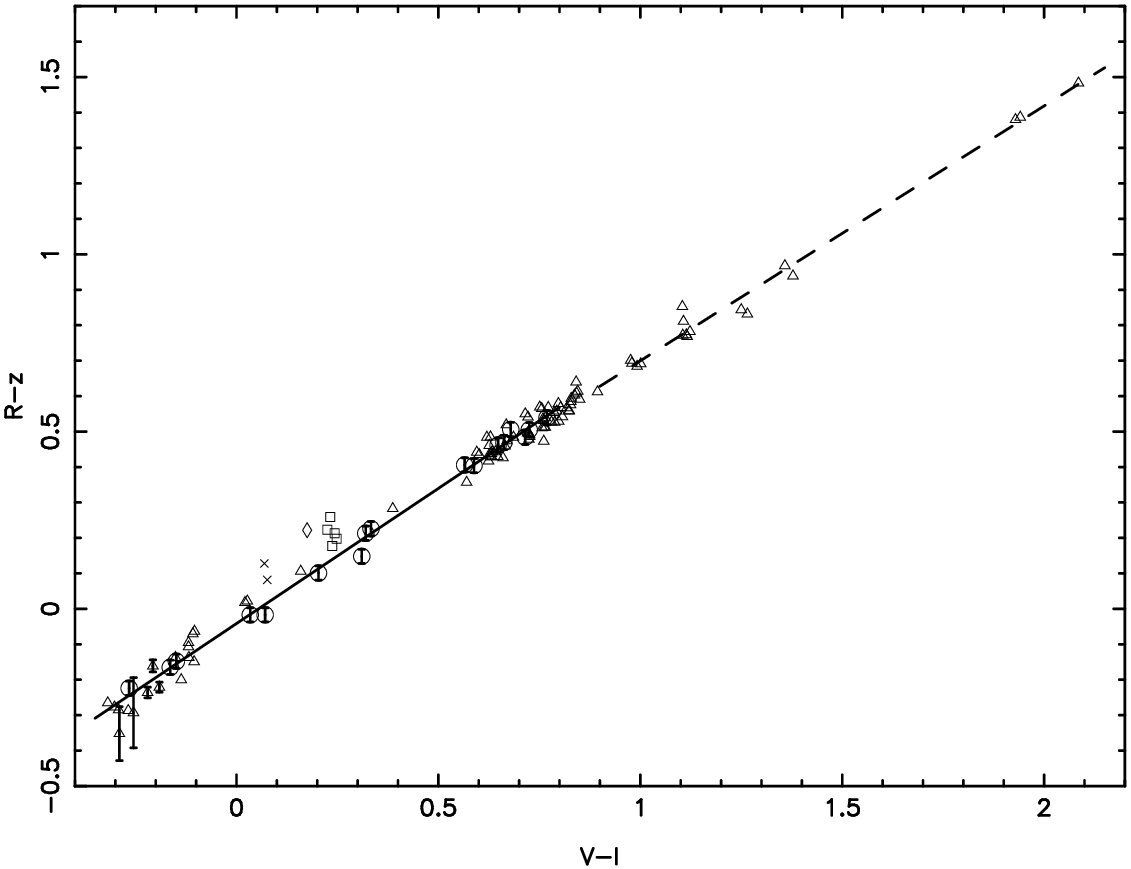] {$R-z$ vs. $V-I$ colors for spectrophotometric
standards (Stone \& Baldwin 1983; open circles) and for Landolt (1992a)
standards.  The $z$ magnitudes of Stone-Baldwin standards were taken from
Hamuy (2001, Appendix B); the solid line is a fit to data for these stars. 
The dashed line is a fit to points with $V-I > 0.97$.  Three particular stars
are Rubin 149C (open squares), SA95-96 (diamond), and Rubin 152E (X's).
See text for further details.  \label{vi_rz}
}

\clearpage

\begin{figure}
\plotone{fig1.eps}
{\center Krisciunas {\it et al.} Fig. \ref{99ee_yjhk}}
\end{figure}

\begin{figure}
\plotone{fig2.ps}
{\center Krisciunas {\it et al.} Fig. \ref{sn2000bh}}
\end{figure}

\begin{figure}
\plotone{fig3.eps}
{\center Krisciunas {\it et al.} Fig. \ref{00bh_opt_ir}}
\end{figure}

\begin{figure}
\plotone{fig4.ps}
{\center Krisciunas {\it et al.} Fig. \ref{sn2000ca}}
\end{figure}

\begin{figure}
\plotone{fig5.eps}
{\center Krisciunas {\it et al.} Fig. \ref{00ca_opt_ir}}
\end{figure}

\begin{figure}
\plotone{fig6.ps}
{\center Krisciunas {\it et al.} Fig. \ref{sn2001ba}}
\end{figure}

\begin{figure}
\plotone{fig7.eps}
{\center Krisciunas {\it et al.} Fig. \ref{01ba_opt_ir}}
\end{figure}

\clearpage
\pagestyle{empty}
\begin{figure}
\plotone{fig8.ps}
\vspace {5 mm} 
{\center Krisciunas {\it et al.} Fig. \ref{j_light_curves}}
\end{figure}

\clearpage
\pagestyle{plain}
\begin{figure}
\plotone{fig9.eps}
{\center Krisciunas {\it et al.} Fig. \ref{jhk_maxima}}
\end{figure}

\begin{figure}
\plotone{fig10.eps}
\vspace {5 mm}
{\center Krisciunas {\it et al.} Fig. \ref{i_stretch}}
\end{figure}

\begin{figure}
\plotone{fig11.eps}
\vspace {5 mm}
{\center Krisciunas {\it et al.} Fig. \ref{vjhk}}
\end{figure}

\begin{figure}
\plotone{fig12.eps}
\vspace {5 mm}
{\center Krisciunas {\it et al.} Fig. \ref{vjhkmax}}
\end{figure}

\begin{figure}
\plotone{fig13.eps}
\vspace {5 mm}
{\center Krisciunas {\it et al.} Fig. \ref{vjh6}}
\end{figure}

\clearpage
\pagestyle{empty}
\begin{figure}
\plotone{fig14.ps}
\vspace {5 mm} 
{\center Krisciunas {\it et al.} Fig. \ref{vi_rz}}
\end{figure}

\end{document}